\documentstyle[epsf]{article}

\def\myhalfypicture#1#2{%
\leavevmode
\unitlength 0.8cm
\epsfverbosetrue
\epsfxsize=281.6pt
\epsfysize=138.22pt
\begin{picture}(12,6)(-1.8,-0.7275)
\put(-1.9545,-0.795){ \epsfbox{#1}}
{#2
}
\end{picture}
}

\def\mytwothirdfpicture#1#2{%
\begin{center}
\leavevmode
\unitlength 0.66666cm
\epsfverbosetrue
\epsfxsize=234.66666667pt
\epsfysize=208.667pt
\begin{picture}(7,12)(1.2,-0.7275)
\put(-2.1545,-0.8){ \epsfbox{#1}}
{#2
}
\end{picture}
\end{center}
}

\advance \textheight by 0.75cm

\def\qed{\hbox{${\vcenter{\vbox{                          
   \hrule height 0.4pt\hbox{\vrule width 0.4pt height 6pt
   \kern5pt\vrule width 0.4pt}\hrule height 0.4pt}}}$}}

\newcommand{\chemrarrow}[1]{
  {\unitlength1ex {\put(0,0.7){\makebox(6,3){$#1$}}
      \makebox[6ex][l]{\put(0,0.75){\vector(1,0){6}}}}}}
\newcommand{\rnd}{\mbox{\,rnd}}
\newcommand{\ld}{\mbox{\,ld}}
\newcommand{\floor}{\mbox{\,floor}}
\begin{document}
\normalsize
\thispagestyle{empty}
\setcounter{page}{1}

\renewcommand{\thefootnote}{\fnsymbol{footnote}} 
\def\bsc{{\sc a\kern-6.4pt\sc a\kern-6.4pt\sc a}}
\def\bflatex{\bf L\kern-.30em\raise.3ex\hbox{\bsc}\kern-.14em
T\kern-.1667em\lower.7ex\hbox{E}\kern-.125em X}


\vspace*{0.88truein}

\centerline{\bf THE MARKOFF--AUTOMATON}
\vspace*{0.035truein}
\centerline{\bf A NEW ALGORITHM FOR SIMULATING THE TIME--EVOLUTION}
\vspace*{0.035truein}
\centerline{\bf OF LARGE STOCHASTIC DYNAMIC SYSTEMS}
\vspace{0.37truein}
\centerline{\footnotesize THOMAS FRICKE\footnote{Email
thomas@summa.physik.hu-berlin.de}}
\vspace*{0.015truein}
\centerline{\footnotesize\it Institut f\"ur Physik}
\vspace*{0.015truein}
\centerline{\footnotesize\it Humboldt--Universit\"at}
\vspace*{0.015truein}
\centerline{\footnotesize\it Invalidenstr.~110, D--100\,99 Berlin, Germany    }
\vglue 10pt
\centerline{\rm and}
\vglue 10pt
\centerline{\footnotesize DIETMAR WENDT\footnote{Email
dietmar@acds16.physik.rwth-aachen.de}}
\vspace*{0.015truein}
\centerline{\footnotesize\it Institut f\"ur Theoretische Physik}
\vspace*{0.015truein}
\centerline{\footnotesize\it Rheinisch Westf\"alische Technische Hochschule
Aachen }
\vspace*{0.015truein}
\centerline{\footnotesize\it Sommerfeldstra\ss e, D--52\,056 Aachen, Germany }
\baselineskip=10pt
\vspace{0.225truein}

\vspace*{0.21truein}
\abstract{We describe a new algorithm for simulating
complex Markoff--processes.
We have used a reaction--cell method
in order to simulate  arbitrary reactions.
It can be used for any kind of  RDS  on arbitrary topologies,
including fractal dimensions or configurations not being related to
any spatial geometry.
The events within a single cell are managed by an event handler
which has been implemented independently of the system studied.
The method is exact on the Markoff level including the correct
treatment of finite numbers of molecules.
\\
To demonstrate its properties,
we apply it on a very
simple reaction--diffusion--systems (RDS).
The chemical equations $\scriptstyle A+A\to \emptyset$ and $\scriptstyle A+B\to
\emptyset$ in
1 to  4 dimensions serve as  models for systems
whose dynamics on an intermediate
time scale are governed by fluctuations.
We compare our results to the analytic approach by
the scaling ansatz.
The simulations confirm the exponents of the $\scriptstyle A+B$ system within
statistical errors, including the logarithmic
corrections in the dimension  $\scriptstyle d=2$.
\\
The method is capable to simulate the crossover from
the reaction to diffusion limited regime, which
is defined be a crossover time depending on the
system size.}{}{}

\vspace*{1pt}

\section{Introduction}
Thus the dynamics of chemical reactions in homogeneous
systems  is well understood,  they  are easy to  describe by
a van't Hoff ansatz  \cite{vk}.
However, in inhomogeneous systems the transport mechanism has to be
taken into account.
We want to consider reaction--diffusion--systems (RDS) for
which the slowest mechanism determines the dynamics of a RDS.
RDS are known to be capable of a very complex behaviour
like formation of spatial and temporal patterns.
They show this great variety of effects,
even if their discrete nature
can be neglected and they  are well described by the
mean concentrations $n_\alpha(\vec r)$.
Population dynamics, which is often described by a formulation equivalent to
chemical reactions, deals with much smaller concentrations, so the
role of fluctuations is much more important than for chemical reactions.

\subsection{The structure of this paper}
At first, to introduce the notation,
we briefly  review the  general method for
simulating master--equations, with respect to chemical reactions.
Then we will present the reaction--cell model to describe extensive
reaction volumes.
After that, we give a detailed description of the implementation
as a the "Markoff--automaton".
Our algorithm makes use of non--numerical data structures borrowed
from computer science,
because we need programming methods
not commonly used in standard numerics.
Therefore, we describe our methods in details.
They  may be regarded as  ''semi--numerical'' algorithms in the sense
of \cite{kn2}.
Finally,  we will present the results we have gained applying
our algorithm  to the simplest
RDS $A+A\to \emptyset$ and $A+B\to \emptyset$ in
1 to 4 dimensions.

\subsection{Reaction--Diffusion--Systems}
We want to denote different types of molecules $\Xi_\alpha$
by a Greek index $\alpha$,
while the reactions are distinguished by the Latin letter $i$.
The stoichiometric equations  describe
the chemical reaction of the $i$--th type unambiguously
by the initial (or forward) $f_\alpha^i $ and final (or backward)
 $b_\alpha^i$ coefficients and the related reaction--rate $\lambda^i$
\begin{equation}
\label{reaction}
  \sum_\alpha f_\alpha^i\, \Xi_\alpha
\chemrarrow{\lambda^i}
  \sum_\alpha b_\alpha^i\, \Xi_\alpha.
\end{equation}

The common formulation of the mean--field equations for
position--dependent mesoscopic concentrations $n(x,t)$ leads to
nonlinear coupled partial differential equations
\begin{equation}
  \frac{\partial }{\partial t}n_\alpha=
  D_\alpha \Delta n_\alpha +\sum_i \lambda^i \prod_\beta
     (b_\beta^i-f_\beta^i)\, n_\beta^{f_\beta^i}.
\end{equation}
In  some realisations of certain systems they may also describe
the fluctuations  \cite{sokol,sokol:al} or give only quantitative changes
\cite{zumofen:al}, nevertheless, it may become necessary
to respect the influence of fluctuations, especially in biological systems,
where the number of individual molecules is very small \cite{zfph90,mydipl}.

However, in contrast to the simple derivations of the partial differential
equations for the mean values,
the formulation of the equations for the higher moments
are much more complicated, even for simple systems.

Note the different scaling of mean concentrations and fluctuations.
While concentrations are  proportional to the number of particles
 $N_\alpha$ in a  volume $\Omega$, i.e~$ N_\alpha=\Omega n_\alpha$,
 fluctuations scale with a different power law according to
$\delta N_\alpha\cong \sqrt{N_\alpha}$.
If we introduce the fluctuations of $n_\alpha$
as  additional quantities $\delta n_\alpha$, we have to take into
account, that they usually scale as  $\delta n_\alpha\cong \sqrt{ N_\alpha }$.
However, the assumption of a $\Omega^{1/2}$ scaling is invalid,
if fluctuations may  increase, as in our case.
Other important examples of scaling departing
from this law are physical systems showing critical behaviour.
A rigourous derivation of the  fluctuation scaling
by terms of an $1/\Omega$  expansion is given in \cite{vk}.

Therefore, the analytical treatment turns out to become very difficult,
if diffusion and the discrete nature of the molecules together
must be taken into account.
Not only the cost of computation grows proportional to the volume,
it has to be remarked that it also increases exponentially to the
statistical moments taken into consideration.

In the reaction--limited case, diffusion is sufficiently fast  to
keep the concentrations homogeneous. Thus the reactions follow
a global van't Hoff dynamics, for which diffusion may be neglected.
For the diffusion--controlled reactions the time to pass structures like
domains or empty spaces determines the time scale of the RDS.
To introduce spatial resolution we subdivide the volume $\Omega$
into $L^d$ small cells of size $\omega=\Omega/L^d$.

Therefore, for small numbers of molecules within an $\omega$ volume,
the discrete nature of the molecules must be taken into consideration.
The raw estimation for systems of $N$  molecules leads to an estimation for
the  fluctuations of $O(\sqrt{N})$.
Thus we need $O(N)=10,000$ molecules
to guarantee a relative error of percentage order if  fluctuations are
neglected.

Former simulations \cite{TousWil,kang:redner,kang:redner2}
using cellular automata or lattice gas methods
to study the dynamics of RDS are restricted to small
volumes and small numbers of molecules.
Computer simulations of complex RDS focus on the
mean--field--behaviour taking no notice of the fluctuations,
whereas on the other hand stochastical simulations of extensive systems
close to equilibrium are not well suited to dynamics.
In contrast, using our algorithm
we may handle  some $10^7$ particles  and are, furthermore, able to
study the crossover from the diffusion--controlled
to the reaction--controlled limit.
Following an analysis of the relevant time scales, the diffusion can be
simulated
on a coarse lattice of cells without affecting the exactness of the results.
Our goal is to close the gap between solving the mean--field--equations
and simulation techniques using cellular automata.
Both techniques  are included, but we want to emphasize,
that standard algorithms may be more efficient to cover these limits.
Because our method  is related to cellular automata,
we call it a "Markoff--automaton" claiming its exactness on the
level of the Markoff--processes derived.

The algorithm has been written for systems with a clear separation of
the short time--scale of reaction events and the long time--scale of
the decay of the population. It should be used with care,
when the scales are mixed, f.e.~by local ordering phenomena.

\subsection{The physical systems}
We consider two different RDS consisting of one of
the two simplest chemical reactions:
\begin{eqnarray}
  A+A&\to& \emptyset,\mbox{\ or \ }\\
  A+B&\to& \emptyset.
\end{eqnarray}
\subsection{The reaction--limited case}
In this case, the diffusion is as fast, that the
extinction is controlled  by the reaction.
Thus  both systems may be described by a global
  van't Hoff ansatz, which leads to the differential equations
for the concentrations $n_\alpha$
\begin{eqnarray}
\label{AAreal}
  \frac{d}{dt}n_A(t)&=&-k\, n_A^2(t),\\
\label{ABreal}
  \frac{d}{dt} n_{A,B}(t)&=&-k\, n_A(t)n_B(t).
\end{eqnarray}
The straightforward solution of (\ref{AAreal}) reads
\begin{equation}
  \label{AAmeanresult}
  n_A(t)= \frac{n_A(0)}{n(0)kt+1}.
\end{equation}
To solve (\ref{ABreal}), we take advantage of the fact, that
the difference  $\Delta n(t)=n_A(t)-n_B(t)=\Delta n(0)$ is  conserved.
This way we obtain the mean--field result
\begin{eqnarray}
  \label{ABmeanresult}
  n_A(t)&=&\frac{n_A(0)\Delta n \exp(-\Delta nkt)}{n_B(0)-n_A(0)\exp(-\Delta
nkt)},\\
  n_B(t)&=&n_A(0)+\Delta n.
\end{eqnarray}
For the case of equal initial concentrations, i.e~$\Delta n=0$,
equation (\ref{ABmeanresult}) reduces to (\ref{AAmeanresult}).
Note that the initial condition $n_A(0)=n_B(0)$ is responsible
for the algebraic decay.

\subsubsection{The diffusion limited case for $A+A\to 0$}
In this case the extinction is controlled
by the speed of diffusion, which  has been analysed in detail
by the
scaling ansatz of Kang and Redner
\cite{kang:redner,kang:redner2}.
However, at this point, we can  only give a brief description of
their main arguments, because we have  outlined their scaling
in the appendixes.

While we want to use this simple model to test our algorithm,
we also  want to examine the accuracy of the scaling ansatz,
which predicts an algebraic  annihilation  for the diffusion
controlled $A+A$ reaction.
The time $t_\xi$ is the time to pass the mean distance between
two $A$ molecules at $t=0$,
\begin{equation}
  n_A(t)\propto  \left\{\begin{array}[h]{lll}
  {t^{-d/2}}, &d<2,\qquad  t >  t_\xi,\\
  {t^{-1}}, &d\geq 2.
\end{array}\right.
\end{equation}
This RDS shows a critical dimension $d_c=2$,
for $d\geq d_c$ the decay is predicted to be determined by mean field
behaviour.

\subsubsection{The diffusion limited case for $A+B\to 0$}
In the $A+B$ reaction  we obtain
the  time scale $t_\xi$ determined by  the time to pass a domain as
\begin{equation}
  t_\xi=\frac{1}{D}\left|\sqrt{n_A(0)}-\sqrt{n_B(0)} \right|^{-{2}/{d}}.
\end{equation}
An algebraic annihilation
 $n_{AB}\propto t^{-\frac{1}{4}}$  with $d_c=4$ appears on
a time scale $t<t_\xi$.
Thus, if $n_A(0)<n_B(0)$ we obtain
in the limit $\Delta n=n_A(0)-n_B(0)\to 0$,
\begin{equation}
  n_{A,B}(t)\propto
  \left\{
  \begin{array}[h]{lll}
    {t^{-d/4}}, &d<4,\qquad t < t_\xi,\\
    {t^{-1}}, &d\geq 4.
  \end{array}
\right.
\end{equation}
A more detailed explanation of the scaling arguments
is left to the appendixes.
Both of these  simple reactions are the most
important examples for the influence  of fluctuations in RDS.
Therefore, to improve our understanding of RDS beyond
mean--field--behaviour, we present our simulation
method as an approach correctly treating
dynamics and statistics in a natural way.

\section{Simulating  master--equations}
We want to  treat a chemical reaction as a Markoff--process,
which is described by the time evolution of the probability $p_{\vec X} $ to be
in state $\vec X$ at time $t$
following a stationary master--equation
\begin{equation}
\label{master}
  \frac{dp_{\vec X}}{dt}=\sum_{\vec X'}
  \underbrace{\left\{W_{{\vec X'}\to{\vec X}}\right.\  p_{\vec X'}}_{
    \parbox[h]{3cm}{\centering inflow into  $\vec X$}}-
  \underbrace{
  \left.W_{{\vec X}\to {\vec X'}}\ p_{\vec X}\right\}}_{
    \parbox[h]{3cm}{\centering outflow  from $\vec X$}},
\end{equation}
in our case the discrete vector $\vec X$
denoting the number of molecules \cite{gi76}.

The minimal--process--method simulates the master--equation as a random--walk
in the space of all possible configurations.
Starting the random--walk  in state $\vec X$,
we only need to determine the  time
 leaving  $\vec X$, and the successor state $\vec X'$,
which is equivalent to draw the lifetime $\tau_{\vec X}$ and
the transition $\vec X\to \vec X'$ as random numbers.

For both steps the knowledge of the flux out of $\vec X$ is completely
sufficient.
The lifetime $\tau_{\vec X} $ is exponentially distributed,
while the selection of the transition requires
the drawing of a random--number  $\vec X'$ proportional to the transition
rate $W_{{\vec X}\to {\vec X'}}$.
This property is due to discrete Markoff--processes, and we want to
emphasize that it is {\it exact} and not affected by
any further assumptions.

To explain the name "minimal--process--method" it has to be
remarked, that the master--equation (\ref{master}) may be interpreted
in different ways.
We may look upon  the outflow term as the  death--process and
upon the flux into a state as the birth--process of the
state $\vec X$ \cite{kt1,kt2}.
The {\it minimal--process} being the process
with the least number of events is
the random--walk--process
in the discrete configuration space of all possible $\vec X$.
The minimal process distinguishes on the rule, that
a death process in $\vec X$ always coincides to a simultaneous
birth process in $\vec X'$.
With the sum
\begin{equation}
  W_{\vec X}=\sum_{\vec X'} W_{{\vec X}\to {\vec X'}},
\end{equation}
the way to generate this random--walk and the sequence of
states $\vec X(t)$ is given by the following simple algorithm:
\begin{center}
\fbox{
\advance\textwidth by -7em
\parbox{\textwidth}{
\noindent{\bf Minimal--process--method}
\begin{enumerate}
\item Generate an exponentially distributed random number $\tau$   with\\
  \mbox{$\Pr(\tau)=W_{\vec X}\cdot \exp(-W_{\vec X}\cdot\tau),\qquad$}
  let $ t\leftarrow t+\tau$,
\item select $\vec X'$ according to the probability \\
   $\Pr(\vec X\to \vec X')={W_{\vec X\to \vec X'}}/{W_{\vec X}}$,
\item go back to step 1.
\end{enumerate}}
\advance\textwidth by 7em
}
\end{center}
The characteristics of discrete processes
were  already known to Markoff, the minimal--process--method
at least since the early sixties \cite{kt1,kt2}.
However, no one has made use of it for computer simulations
until the mid seventies when Gillespie applied it to chemical
reaction--systems \cite{gi76,gi78}.

\subsection{Chemical reactions in a homogeneous volume}
For the present, we want to discuss arbitrary chemical reactions in
a small volume $\omega$, for which the diffusion is so fast
that any spatial inhomogeneity may be neglected.
After that, we will introduce diffusion as a reaction--like process
of molecules leaving the volume $\omega$.
In this way, we want to describe the dynamics by which
diffusion relates
distances of an extensive length scale to a time scale.

According to (\ref{reaction})
the number  $X_\alpha$ of molecules $\Xi_\alpha$
changes
from  $X_\alpha$ to $ X_\alpha + b_\alpha^i - f_\alpha^i$
each time the reaction of the $i$--th type occurs.
I.e.~$f_\alpha^i$ molecules $\Xi_\alpha$ are consumed
and  $b_\alpha^i$ molecules $\Xi_\alpha$ are produced.
Forward and backward reactions shall be distinguished by
their enumerations. If the reactions $j$ and $i$
are reciprocals   of each other, the role of the coefficients is
exchanged, i.~e. $ f_\alpha^i  = b_\alpha^j,$ and $ f_\alpha^j = b_\alpha^i$.

Since the number of molecules is an extensive quantity,
the number of chemical reactions per unit time must  be
extensive, too.
The  rate of events, the  reactivity $\Lambda^i$ of a chemical reaction
is defined as
\begin{equation}
  \Lambda^i dt=\Pr\left\{\parbox{30ex}{the reaction $i$
occurs during the time interval $[t,t+dt]$}\right\}.
\end{equation}
According to van't Hoff \cite{vk}, it is given by
\begin{equation}
\label{vantHoff}
  \Lambda^i=\lambda^i \omega \prod_\alpha  \frac{X_\alpha
!}{(X_\alpha-f_\alpha^i)!\
    \omega^{f_\alpha^i}},
\end{equation}
where $\lambda^i$ is an intensive constant which does not depend on
$X_\alpha $.
Note that  the  reactivities $\Lambda$
depend on the state,
i.e.~$\Lambda^i=\Lambda^i_{\vec X},\ \Lambda=\Lambda_{\vec X}$,
the index ${\vec X}$ is usually  suppressed.
The fraction
\begin{equation}
  \frac{X_\alpha !}{(X_\alpha-f_\alpha^i)!\ \omega^{f_\alpha^i} }=
  \frac{X_\alpha(X_\alpha-1)\cdots
(X_\alpha-f_\alpha^i+1)}{\omega^{f_\alpha^i}}
\end{equation}
takes into consideration that each individual molecule can only be consumed
once. This is important, if the numbers $X_\alpha$ are small.
For large $X_\alpha$, this fraction turns into the more familiar
form
 $  \left(\frac{X_\alpha }{\omega}\right)^{f_\alpha^i}$,
ignoring any power up to
$  \left(\frac{X_\alpha }{\omega}\right)^{f_\alpha^i-1}$.

The van't Hoff approach suggests chemical
reactions being Markoff--processes \cite{gi76,gi78},
which may be justified due to a stosszahlansatz for homogeneous
concentrations.
The configuration  is the vector consisting of the number of
molecules
\begin{equation}
  {\vec X}=(X_1,X_2,\ldots,X_\alpha,\ldots).
\end{equation}
For abbreviation we  group the initial and final coefficients for
the reaction of the $i$--th type  as vectors
 $ {\vec f}^i=(f_1^i,f_2^i,\ldots,f_\alpha^i,\ldots),
 \quad {\vec b}^i=(b_1^i,b_2^i,\ldots,b_\alpha^i,\ldots)$.
Each type of reaction is realized by an integer vector
operation, the vector $\vec \delta^i={\vec b}^i- {\vec f}^i$ denoting
the changes caused by the $i$--th reaction:
\begin{equation}
\label{reaction:delta}
  {\vec X}\leftarrow{\vec X}+ {\vec \delta}^i.
\end{equation}
In our case the vector $\vec X$ is simply a one-- or two-dimensional
vector
\begin{equation}
  \vec X=\left\{ \begin{array}[h]{rl}
  (X_A)&\mbox{ reaction $A+A\to \emptyset$},\\
  (X_A,X_B)&\mbox{ reaction $A+B\to \emptyset$}.
\end{array}\right.
\end{equation}
The reaction--rates are computed  according (\ref{vantHoff}) as
\begin{eqnarray}
  \Lambda_{AA}&=&\lambda_{AA}\omega \frac{X_A(X_A-1)}{\omega^2},\\
  \Lambda_{AB}&=&\lambda_{AB}\omega \frac{X_A X_B}{\omega^2}.
\end{eqnarray}
The homogeneity of the concentrations within $\omega$ is necessary
for  the van't Hoff ansatz. We will have to
subdivide the volume into sufficiently small cells to justify a local form
of (\ref{vantHoff}), introducing diffusion as hopping from one
cell to an adjacent one.

\subsection{Gillespie's algorithm}
With these assumptions
Gillespie describes the sequence of states
of a  chemical system by the random--walk of the vector
${\vec X}={\vec X}(t)$.
His algorithm is a version of the minimal--process--method adapted
to the dynamics of chemical reactions, which is based on the well
known properties of discrete stochastic processes\cite{kt1,kt2,gi76,gi78}.
We summarise his main results according to our requirements.
The total reactivity is the rate
\begin{equation}
  \Lambda=\sum_{\mbox{all reactions $i$}} \Lambda^i,
\end{equation}
which determines the probability of a chemical reaction during the
infinitesimal
time interval $[t,t+dt]$ according to
\begin{equation}
  \Lambda dt=\Pr\left\{\parbox{40ex}{some
    chemical reaction changes the state $\vec X$ during $[t,t+dt]$}\right\}.
\end{equation}
 $\Lambda$ is the inverse of the average lifetime of $\vec X$,
therefore the random time $\tau_\Lambda$ until the next event
occuring  has the probability density
\begin{equation}
  \Pr\{\tau_\Lambda,\mbox{time to next reaction }\}=\Lambda
\exp(-\Lambda\tau_\Lambda).
\end{equation}
Because we have assumed that  chemical reactions are  Markoff--processes,
this equation yields independently of the time when
the last change  occurred, which
 is a general statement for any discrete Markoff--process.
According to Gillespie, the probability of the
reaction of kind $i$ is proportional to its contribution to $\Lambda$,
thus
\begin{equation}
  \Pr\{\mbox{Reaction }i\}=\frac{\Lambda^i}{\Lambda}
\end{equation}
is the probability, that the next change of $\vec X$ in the chemical system
is determined by a reaction of type $i$ defined by (\ref{reaction:delta}).
The transition rate $W_{{\vec X}\to{\vec X'}}$ is given
by the probability of all chemical reactions
leading from ${\vec X}$ to ${\vec X'}$
\begin{equation}
\label{transition}
  W_{{\vec X}\to{\vec X'}}=
  \sum_{\parbox[h]{20ex}{\centering\small all reactions $i$ with \\
      $ {\vec X'}={\vec X}-{\vec f}^i+{\vec b}^i$}} \Lambda^i.
\end{equation}
Note that different chemical reactions may change $\vec X$ by the same
  $\vec \delta^i=\vec f^i - \vec b^i$ because only
the backward and forward stoichiometric coefficients
  $\vec f^i $ and $ \vec b^i$ together
determine the reaction unambiguously.

Gillespie's algorithm implicitly uses equation (\ref{transition}) to
decompose the
relation of reaction and transition probabilities.
For arbitrary chemical reactions it
computes  the random  time-step $\tau_\Lambda$ until the next reaction by
drawing an uniformly distributed random number \mbox{$\rnd\in [0,1)$}.
The reaction type $i$ usually is selected by a simple loop, known as
linear selection algorithm using the sum $s$ for the integration.
The variable $t$ denotes the time, proceeding in
exponential time steps. The order of reactions denoted by $i$
is arbitrary.
After the reaction has been carried out,
the reaction reactivities $\Lambda^i$ and the total rate
$\Lambda=W_{\vec X}$ are computed again.
\begin{center}
  \fbox{\advance \textwidth by -7em
    \parbox{\textwidth}{
      \noindent {\bf Gillespie's algorithm}
      \begin{enumerate}
      \item $  \tau_\Lambda \leftarrow -\frac{1}{\Lambda}\log(1-\rnd),\qquad
        t\leftarrow t+\tau_\Lambda$
      \item $r\leftarrow \rnd, \quad s\leftarrow 0,\quad i'\leftarrow$first
reaction,
      \item \parbox[t]{10em}{while $s \leq r \Lambda $,}
       \fbox{\parbox[t]{20em}{
          \parbox{3em}{(a)} $i\leftarrow i', \quad i'\leftarrow$next reaction\\
          \parbox{3em}{(b)} add up $s\leftarrow s+\Lambda^i$}}
       \item  do reaction type $i$ by  ${\vec X}\leftarrow{\vec X}+{\vec
\delta}^i$,
        compute $\Lambda^i,$ and $ \Lambda\leftarrow \sum_i \Lambda^i$,
      \item go back to step 1.
      \end{enumerate}
      \advance \textwidth by 7em}}
\end{center}
Note the adaptation to the time scale $\Lambda^{-1}$,
which is responsible for  the high flexibility of
Gillespie's algorithm.
Thus it is a fast and easy to use Monte--Carlo
method for the dynamics of  arbitrary reaction--systems
and has the main advantage over the mean--field description,
that it  takes into account
internal fluctuations caused by the finite numbers of
molecules.
Because the random--walk imitates the sequence of states
of the simulated Markoff--process, any
correlations  and higher moments
may be measured as in a real experiment.

\section{The reaction--cell method}
The molecules are treated as point--like particles
with a finite interaction probability,
implicicelty assuming, that their Brownian motion is independent.
Our simulation introduces a reaction--cell method.
In general, for small diffusion--constants,
the homogeneity condition of the van't Hoff ansatz cannot be
fulfilled for the total $d$--dimensional volume $\Omega$.
Thus $\Omega$ is subdivided into
small cells (cubes or squares etc.) of volume $\omega=\frac{\Omega}{L^d}=h^d$.
We have to impose the condition, that the size of $\omega$ may be justified
by a local van't Hoff ansatz.
This prerequisite depends on the time--scale $\tau_R$ between two subsequent
chemical reactions compared to the time scale $\tau_D$ of a molecule
leaving $\omega$ by diffusion.
The diffusion--time--scale has to be much faster than the reaction
time--scale, i.~e.~$\tau_D\ll \tau_R$ with $\tau_D=h^2/D$
and $\tau_R=1/\Lambda$.

Our algorithm covers the field between the deterministic description of
RDS, which does not describe the fluctuations correctly,
and the cellular automaton approach, which is not well conditioned
to describe large numbers of particles.
The assumptions have to be justified in any case, and have to be
modified carefully, f.e.~if
the spatial extension of the moving objects has to be taken into account.
This is important for the $A+A\to 0$ reaction in lower dimensions
\cite{blumen:al,kuzovkov:kotomin},
where the seed reaction constant
may become a irrelevant quantity.
In this case first principle Monte--Carlo simulations have to be
performed and our algorithm has to be checked.

\subsection{Reaction--Diffusion--Systems}
We distinguish different cells  by an integer vector index $\vec n$.
Therefore,
 the possible reactions now depends on the local situation which is described
by the number $X_{\alpha\vec n}$  of molecules $\Xi_\alpha$ in cell $\vec n$.
The total volume $\Omega$ is represented by
the direct sum
\begin{equation}
  \widetilde X=\bigoplus_{\vec n}{\vec X}_{\vec n}=
  \bigoplus_{\vec n }\bigoplus_{\alpha}  X_{\alpha\vec n},
\end{equation}
consisting of  sub-vectors
 $\vec X_{\vec n}=\left\{\ldots, X_{\alpha\vec n},\ldots \right\} $ of the
numbers
of molecules in each cell.

For the $A+A\to \emptyset$ system
$\widetilde X$ has the form
\begin{equation}
  \widetilde X =
  \left\{
    \left(X_{A(1,\ldots, 1)}\right)\right.,
      \left(X_{A(1,\ldots, 2)}\right),
        \ldots ,
        \left.\left(X_{A(L,\ldots, L)}
        \right)
      \right\},
    \end{equation}
respectively for the $A+B\to \emptyset$ reaction
\begin{eqnarray}
  \widetilde X =
  \left\{
  \left(X_{A(1,\ldots, 1)},X_{B(1,\ldots, 1)}\right)\right.,
  \left(X_{A(1,\ldots, 2)}\right.,
    \left. X_{B(1,\ldots, 2)}\right),&\\\nonumber
      &\hspace*{-3cm}\ldots,\left.\left(X_{A(L,\ldots, L)},X_{B(L,\ldots,
L)}\right)
\right\}
\end{eqnarray}
In this section we want to introduce diffusion as a random--walk
on the lattice of reaction--cells.
The diffusion in an arbitrary, not necessarily euclidian, topology may be
regarded  as a reaction--like step of a molecule $\Xi_\alpha$
from $\vec n$ to one of its
next neighbours $\vec m$,
\begin{equation}
  \Xi_{\alpha \vec n}\qquad
  \chemrarrow{\lambda_{\alpha\vec n\vec m}}\qquad  \Xi_{\alpha \vec m}.
\end{equation}
For a diffusion--step of a molecule $\Xi_{\alpha}$
 the rate
\begin{equation}
  \lambda_{\alpha\vec n\vec m}=D_{\alpha\vec n\vec m}\frac{ 1}{h^2},
\end{equation}
with $D_{\alpha\vec n\vec m}$ denoting the local diffusion--constant,
describes  the probability
\begin{equation}
  \lambda_{\alpha\vec n\vec m}dt=\Pr\left\{\parbox{40ex}{a single molecule
    $\Xi_\alpha$ jumps from $\vec n$ to $\vec m$ during the time intervals
    $[t,t+dt]$}\right\}.
\end{equation}
On a lattice built of volumes  $\omega=h^d$, we obtain
in the limit $t \to \infty$ the correct Brownian motion
behaviour by the central--limit--theorem.
The symbol  $\langle\vec n\rangle $ denotes the set of all $2d$
next neighbour cells of $\vec n$.
Due to  the symmetry of the rate to do a single step into any direction
i.e.~the rate of leaving $\vec n$ by diffusion,
 is $D_{\alpha\vec n}/{h^2}$, with
\begin{equation}
   \lambda_{\alpha\vec n}=\sum_{\vec m\in \langle\vec n\rangle }
   \lambda_{\alpha\vec n\vec m}=2d\lambda_{\alpha\vec n\vec m}.
\end{equation}
In the isotropic and homogeneous case we omit the spatial indices
 $  \lambda_{\alpha}=\lambda_{\alpha\vec n}$
and the rate for   stepping   in  a certain dimension gets
 $\lambda_\alpha/2d$, while the total rate for leaving a cell
is $\lambda_\alpha$.

If we have   $X_{\alpha,\vec n}$ molecules $\Xi_\alpha$ in cell $\vec n$,
the  {rate for the next diffusion step of any $\Xi_\alpha$} to $\vec m$ is
\begin{equation}
\label{diffusion}
  \Lambda_{\vec n\vec m}=
  \lambda_{\vec n\vec m} X_{\alpha\vec n}.
\end{equation}
This rate is the equivalent formulation to (\ref{vantHoff}), treating
a diffusion--step like a first order chemical reaction.
If a diffusion--step occurs, another cell $\vec m$ gets involved,
because one $\Xi_\alpha$ moves from $\vec n$ to $\vec m$.
Thus a diffusion--step is carried out by
\begin{equation}
1.\ X_{\alpha\vec n}\leftarrow X_{\alpha\vec n}-1,\qquad
2.\ X_{\alpha\vec m}\leftarrow X_{\alpha\vec m}+1.
\end{equation}
Notice that in our case $\Lambda_{\alpha \vec n\vec m}$
 depends  only on the contents of
 $\vec n$,
from which a molecule hops into a neighbour $\vec m\in \langle \vec n\rangle $,
although the assigned diffusion--step  changes the contents
of the other cell $\vec m$, too.
Thus a diffusion--step changes the state of two cells, however,
its rate depends only on the contents of the first cell $\vec n$
triggering the event.

For the chemical reaction of  type $i$  we   simply have to change the contents
of  $\vec n$ by ${\vec X}_{\vec n}\leftarrow{\vec X}_{\vec n}+{\vec \delta}^i$.
Therefore, the probability $Q_{\vec n}$ of an event triggered by cell
 $\vec n$ is the sum of reaction--rates  $\Lambda^i_{\vec n}$
and diffusion--rates  $\Lambda_{\alpha\vec n\vec m}$
\begin{equation}
\label{Q:rate}
  Q_{\vec n}=
  \sum_i \Lambda^i_{\vec n}+
  \sum_{\alpha}\sum_{\vec m\in \langle\vec n\rangle } \Lambda_{\alpha\vec n\vec
m}
\end{equation}
Because all reaction-- and diffusion--steps are assigned to the
cell $\vec n$,
we may identify an arbitrary step entirely by an integer $i$
and the cell $\vec n$.
We may ignore the additional index $\vec m$
for the selection mechanism.
A  reaction may be defined by
an addition  of its associated difference \mbox{$ \widetilde \delta^i_{\vec
n}$},
i.e.~$   \widetilde X\leftarrow{\widetilde X}+ \widetilde \delta^i_{\vec n} $,
involving one or two  cells.
In section 6.1.1 below
we suggest obvious generalisations of rates depending on
the contents of the  neighbourhood of $\vec n$.
This way it is easy to  introduce dynamics depending on gradients,
 like the movement of kinks and steps on surfaces.

\subsection{Times scales of reaction--  and diffusion--processes}
If we want to fulfill the assumptions of a local van't Hoff ansatz we
have to satisfy the condition, that diffusion is  much faster than
any chemical reaction.
More precisely spoken, this means, that the local reactions  $\Lambda_{\vec
n}^i $
and diffusion-- $\Lambda_{\alpha\vec n\vec m}$
rates define different time scales,  which have  to be separated.
 Therefore we choose the size of our cells sufficiently small,
for  the condition
\begin{equation}
\label{timescale}
  \Lambda_{\vec n}^i\ll \Lambda_{\alpha\vec n\vec m},
  \quad \mbox{for any $i,\alpha$}
\end{equation}
to be satisfied.
This can always be achieved, because reaction probabilities
are extensive quantities $\Lambda^i_{\vec n}\propto \omega=h^d$,
while  diffusion--rates increase with $h^{-2}$.
Thus $\lambda_{\alpha\vec n\vec m}\propto \omega^{-{2}/{d}}$ and
 \mbox{$\Lambda_{\alpha\vec n\vec m}\propto \omega^{1-{2}/{d}}$},
the lattice constant $h$ always can be reduced to make the ratio
\begin{equation}
  \frac{\Lambda^i_{\vec n}}{\Lambda_{\alpha\vec n\vec m}}\propto
\omega^{{2}/{d}}=h^{2}
\end{equation}
arbitrarily small.
For any practical purpose it is not possible to give an a priori estimation,
so we recommend to choose $\omega$ as large as possible   rejecting the
simulation runs if  $10 \Lambda^i_{\vec n} > \Lambda^j_{\vec n\vec m}$.

\section{Selecting a single cell by the method of logarithmic classes}
Although we have argued that  there is no principal difference
between  the Gillespie algorithm
in a homogeneous volume and  the method on a lattice of reaction cells,
a significant  complication arises as  the number of possible changes in
each Markoff--step is an extensive quantity.
If we make a straightforward generalisation of Gillespie's algorithm
using a linear selection strategy we run into the problem
of selecting a single event from an extensive quantity of
transitions. The total reactivity
\begin{equation}
  Q=\sum_{\vec n}Q_{\vec n}=\sum_{\vec n} \sum_{i} Q_{\vec n}^i
\end{equation}
again determines the mean time step.
According to condition (\ref{timescale}) we choose the sequence of
transitions considering at first
the more probable diffusion--steps,
afterwards the  possibility of  chemical reactions.
\begin{center}
\fbox{\advance \textwidth by -7em
\parbox{\textwidth}{\noindent{\bf Gillespie's algorithm for reaction--cells
with linear
selection}
\begin{enumerate}
\item $  \tau_Q \leftarrow -\frac{1}{Q}\log(1-\rnd),\qquad  t\leftarrow
t+\tau_Q$,
\item $r\leftarrow \rnd, \quad s\leftarrow 0$,
\item while $s <r Q$, for all cells $\vec n$ \label{stepn},
  \fbox{
    \advance \textwidth by -10ex
    \parbox[t]{12em}{\noindent\begin{enumerate}
    \item next $\vec n$,
    \item while $s <r Q$,\\
      \fbox{\advance \textwidth by -10ex
        \parbox{9em}{\noindent    \begin{enumerate}
        \item next $i$,
        \item sum $s\leftarrow s+Q^i_{\vec n}$,
        \end{enumerate}}
      \advance \textwidth by 10ex
      }
  \end{enumerate}}
\advance \textwidth by 10ex}
\item do transition $\widetilde X\leftarrow\widetilde X+
\widetilde\delta^i_{\vec n}$,
\item for all cells   $\vec n$ and reactions $i$ involved,\\
  recompute all  $Q^i_{\vec n}\Rightarrow Q_{\vec n}\Rightarrow Q $,
\item go back to step 1.
\end{enumerate}
\advance \textwidth by 7em}}
\end{center}

\begin{figure}[p]
  \begin{center}
    \leavevmode
    \myhalfypicture{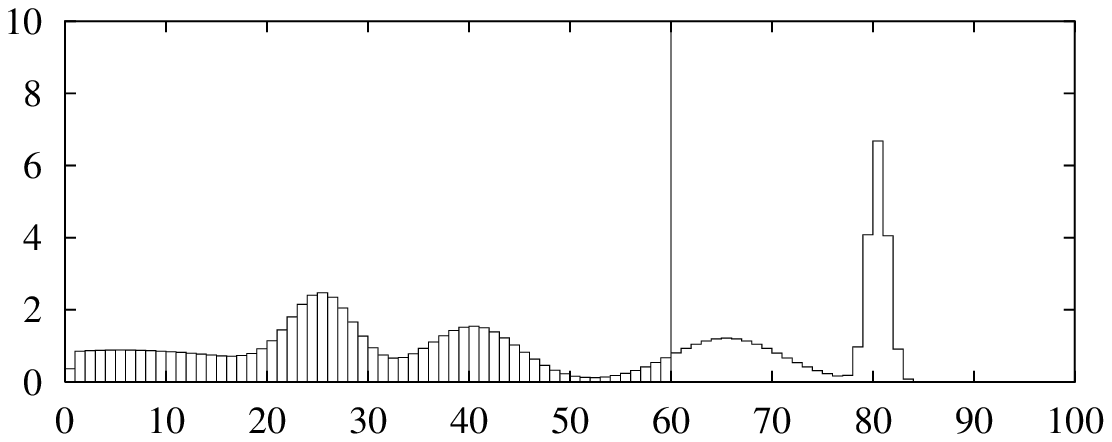}{
      \put(0,-1){\parbox{8cm}{\centering cells $\vec n$}}
      \put(0,5.1){\parbox{8cm}{\centering linear selection}}
      \put(-3.1,5){(a)}
      \put(-1.3,2.5){\centering$Q_{\vec n}$}
      \put(4.7,0.45){\normalsize $s=rQ$}
      \put(1,2.2){$C_{\mbox{\tiny LS}}\propto V$}
      }\\[7mm]
    \myhalfypicture{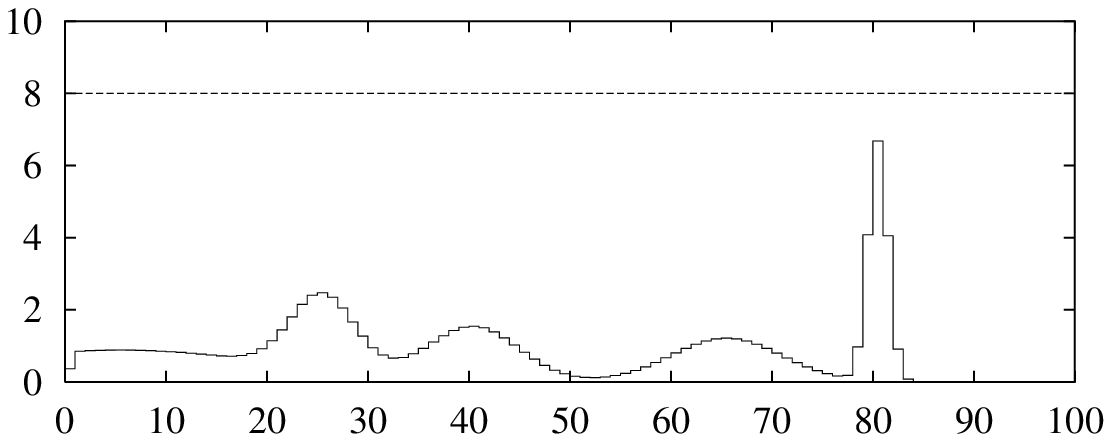}{
      \put(-3.1,5){ (b)}
      \put(5,3.5){$\hat Q$}
      \put(4,1.5){rejected}
      \put(2.5,0.075){\small accepted}
      \put(0,-1){\parbox{8cm}{\centering cells $\vec n$}}
      \put(0,5.1){\parbox{8cm}{\centering  von Neumann rejection}}
      \put(1,2.2){$C_{\mbox{\tiny vNR}}\propto=
        \frac{{\hat Q}\sum_{\vec n}1}{\sum_{\vec n} Q_{\vec n}}$}
      \put(-1.3,2.5){\centering$Q_{\vec n}$}}\\[7mm]
    \myhalfypicture{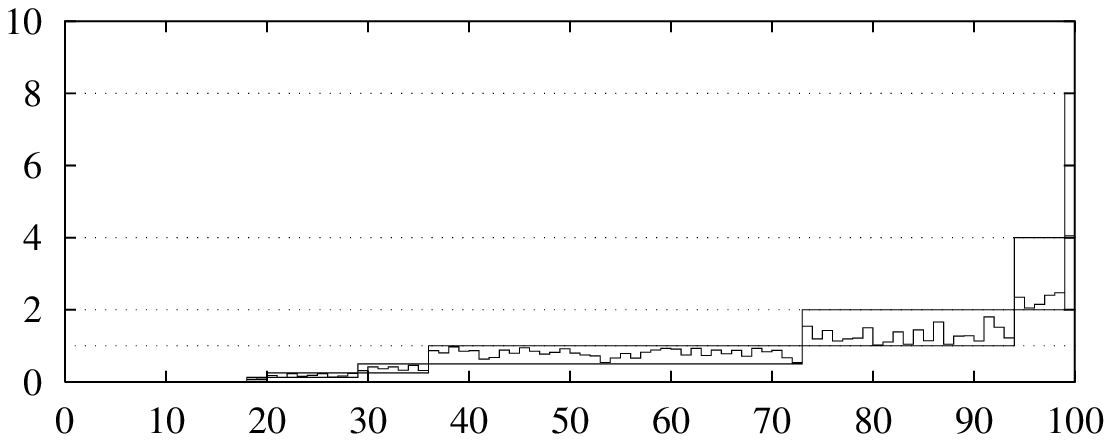}{
      \put(-3.1,5){(c)}
      \put(10.1,-0.1){\parbox[h]{0.5cm}{\hfill $-\infty$}}
      \put(11.5,2.5){$z$}
      \put(10.1,0.40){\parbox[h]{0.5cm}{\hfill $0$}}
      \put(10.1,0.90){\parbox[h]{0.5cm}{\hfill $1$}}
      \put(10.1,1.90){\parbox[h]{0.5cm}{\hfill $2$}}
      \put(10.1,3.90){\parbox[h]{0.5cm}{\hfill $3$}}
      \put(0,-1){\parbox{8cm}{\centering cells $\vec n$, rearranged}}
      \put(0,5.1){\parbox{8cm}{\centering method of logarithmic classes}}
      \put(-1.3,2.5){\centering$Q_{\vec n}$}
      \put(1,2.2){$C_{\mbox{\tiny L2}}=\mbox{constant}$}
      }
\caption{\label{fig:selections}{Selection algorithms:
    (a) Linear s., the loop integrates $Q_{\vec n}$ up to $rQ$,
      thus the cost is extensive,
    (b) Von Neumann rejection without improvement,
      the  cost is the ratio of the area of the
      rectangle marked off by the upper limit $\hat Q$ to the area
      below $Q_{\vec n}$.
      A single peak may severely affect  its efficiency,
    (c) s.~by the method of logarithmic classes, the upper and lower
      limits for each class are
      denoted by dotted lines. The acceptance $a\geq 0.5$ is
      the ratio of the reordered  $Q_{\vec n}$ to its upper estimation
      $2^{\mbox{$ \floor\left(\ld(Q_{\vec n}) \right)+1$}} $,
      we find the cost being approximately constant. }}

\end{center}
\end{figure}

This method suffers from the severe disadvantage
that the sum  $S$ computed at step (\ref{stepn})
adds up an extensive number of reaction probabilities.
Drawing a random cell due to its contribution $Q_{\vec n}$ to $Q$,
the loop  consumes computing
time proportional to the number of cells.
This procedure is completely unsuitable for a single step
of a computer algorithm, see figure \ref{fig:selections}(a).
\advance\textwidth by -7em
\advance\textwidth by 7em

If we do not have any additional information,
we have on the  average  to walk the half length of the main loop.
Even for a small lattice of $10^4=100 \times 100 $ cells this means
an increase for the computing time compared to the algorithm without spatial
resolution by an average   factor  of $5,000$.
It seems to be impossible to circumvent this problem by  rearranging  the cells
or some kind of pre-sorting, because the selected reaction
changes the cell--reactivity every time.
A random selection according A.~J.~Walker's \cite{kn2} algorithm
would be efficient only if the reaction--rates did not depend on time.
The extensive quantity of cells to be considered is a principle obstacle.

These reflections show that  linear selection is totally inefficient.
Thus  we have tried another algorithm based on the von Neumann rejection,
 which selects at first  the
reaction--cell and in a second step the reaction or
diffusion in this cell.

\subsection{Von Neumann rejection}
The von Neumann rejection requires an upper bound for the
probability to select a cell.
 Therefore we have to make the assumption, that
\begin{equation}
  Q_{\vec n}\leq \hat Q \qquad \mbox{for each cell $\vec n$}.
\end{equation}
We replace step (\ref{stepn}) by
\begin{center}
\fbox{\advance \textwidth by -7em
\parbox{\textwidth}{
\begin{enumerate}
\item[(\ref{stepn})] von Neumann rejection\\
\fbox{\advance \textwidth by -10ex
\parbox{\textwidth}{
\begin{enumerate}
  \item select a cell $\vec n$  with uniform probability,
  \item choose a uniformly distributed random number $r\leftarrow\rnd$,
  \item if $ Q_{\vec n}\geq r\hat Q \qquad$
    \advance \textwidth by -22ex
      \parbox[t]{\textwidth}{then
        select a reaction according to $X_{\alpha \vec n}$\\
        else reject this choice and go back to \mbox{step (a).}}
      \advance \textwidth by 22ex
  \end{enumerate}}
\advance \textwidth by 10ex}
\end{enumerate}}
\advance \textwidth by 7em}
\end{center}

On the first glance this algorithm seems to solve all problems.
The method does not depend on the number of cells,
thus the problem of the dependency on the extensive number of cells
does not arise.
However, a more precise look reveals that
the efficiency of the von Neumann rejection
strongly depends on the homogeneity of the RDS.
This is demonstrated in  figure \ref{fig:selections}(b).
The acceptance ratio $a$ is the quotient of the area below
 $Q_{\vec n}$ and the rectangle delimited by $\hat Q$
\begin{equation}
  a=\frac{\sum_{\vec n} Q_{\vec n}}{{\hat Q}\sum_{\vec n}1},
\end{equation}
and $1/a$ is proportional to the number of runs  through the loop needed
to select a cell.
Therefore the cost $C_{\mbox{\tiny vNR}}$ of the algorithm is proportional to
the inverse of the acceptance ratio,
\begin{equation}
  C_{\mbox{\tiny vNR}}\propto \frac{1}{a}
  =\frac{{\hat Q}\sum_{\vec n}1}{\sum_{\vec n} Q_{\vec n}}.
\end{equation}

For very inhomogeneous systems, i.e.~for systems dominated by a single
peak, this method may slow down by an arbitrary factor,
which has been studied for the simulation of a biological system \cite{zfph90}.
This disaster can be avoided,
if we split  up the cells into classes according to their dual order of
magnitude.
The acceptance ratio is improved to 75\%.

\subsection{Method of logarithmic classes}
If we want to  handle all orders of magnitude of the
cell reactivities  $Q_{\vec n} $ we have to implement
a logarithmic classification scheme.
We define the logarithmic class $L_z$
as the set of all cells $\vec n$
with a reactivity $Q_{\vec n}$ in the same order of dual magnitude.
The symbol $\ld(x)=\log_2(x)$ denotes the logarithm with base 2,
and the  function $\floor(x)$ denotes the largest integer which is not
greater than $x$
\begin{equation}
  L_z=\left\{{\vec n} \right|\left. \floor(\ld(Q_{\vec n}))=z
\right\}.
\end{equation}
The reactivity $\ell_z$ of a certain class is given
by the sum of the reactivities of its elements
\begin{equation}
  \ell_z=\sum_{\vec  n\in L_z} Q_{\vec n}.
\end{equation}
There is no principal restriction of the range of $z$
\begin{equation}
  z=\floor(\ld(Q_{\vec n})) \in \{-\infty,\ldots, -2,-1,0,1,2,3,\ldots\}.
\end{equation}
The class $L_{-\infty}$ contains all cells without any possibility
of  a reaction-- or diffusion--step, above all the empty cells.
Because $L_{-\infty}$ cannot be selected, i.e~$\ell_{-\infty}=0$,
there is no need to represent it in computer memory.
For any practical application, $z$ is
restricted to a finite range $z_{\min}\leq z\leq z_{\max}$.
The intersection of two different classes is empty
 \mbox{$L_z\cap L_{z'}=\emptyset$,}\hspace{2em}\mbox{$z\neq z' $,} thus  the
total reactivity is expressed by the sum
\begin{equation}
  Q=\sum_{\vec n}Q_{\vec n}=\sum_{z=-\infty}^\infty \ell_z
  =\sum_{z=z_{\min}}^{z_{\max}} \ell_z.
\end{equation}
The inequality
\begin{equation}
\label{the:inequality}
  2^z\leq Q_{\vec n} < 2^{z+1}, \qquad {\vec n}\in L_z
\end{equation}
guarantees, that the reaction--rates are relatively homogeneous
within a certain class.
The rates of two members of any class  do not differ by more than a factor of
$2$
\begin{equation}
  \frac{\displaystyle \max_{\vec n\in L_z }Q_{\vec n}}{\displaystyle
    \min_{\vec n'\in L_z }Q_{\vec n'}}<2.
\end{equation}
This is an ideal starting point for the von  Neumann rejection.
 We therefore decided to implement an event handler,
which is able to select a reaction--cell according to its reaction--rate
by a von Neumann rejection step within the class.
The number of classes is small, i.e.~ of $O(10)$, thus the choice  of a class
is
based on a linear selection algorithm.

\subsubsection{Selecting a reaction}
The problem has been split into three qualitatively different steps:
\begin{enumerate}
\item  select a class $L_{z}$ with  probability $ {\ell_z}/{Q}$,
\item  select a cell $\vec n\in L_z$ with  probability ${Q_{\vec n}}/{\ell_z}$,
\item  select a reaction--step $i$ within cell $\vec n$ with  probability
 $  {Q^i_{\vec n}}/{Q_{\vec n}}$.
\end{enumerate}
We stress again, that the index $i$ represents both reaction and diffusion
transitions.
The last two steps are selected according their  conditional probabilities
\begin{eqnarray}
\Pr\{\mbox{select class $z$}\}&=& \frac{\ell_z}{Q},\\
\Pr\{\mbox{select cell $\vec n$ $\mid$ class $z$ has been selected}\}
&=& \frac{Q_{\vec n}}{\ell_z},\\
\Pr\{\mbox{select reaction  $i$ $\mid$ cell  $\vec n$ has been selected}\}
&=& \frac{Q^i_{\vec n}}{Q_{\vec n}},
\end{eqnarray}
therefore the probability of selecting a reaction in a cell
\begin{equation}
  \Pr\{\mbox{select reaction  $i$ in cell  $\vec n$}\}=
  \frac{\ell_z}{Q}
  \frac{Q_{\vec n}}{\ell_z}
  \frac{Q_{\vec n}^i}{Q_{\vec n}}
  =\frac{Q_{\vec n}^i}{Q}
\end{equation}
has been maintained correctly.

The algorithm choosing a class is a linear selection.
However, it is not necessary to order the sequence
of classes $(z_i,\ldots,z_j)$
in a naive way with $z_i<z_{i+1}$ or $z_i>z_{i+1}$ for all $i$.
Moreover, it has a favourable effect, if
the classes are sorted   with respect to their reactivity,
i.e~$\ell_{z_i}\geq \ell_{z_{i+1}}$.
To speed up the selection,
the classes of the highest probabilities to be selected are
successively moved to the head of the loop by a
bubble--sort,
see step (\ref{bubble}) below.
This is the most efficient algorithm to select a reaction
in a class according to its rate $\ell_{z_i}$,
because the classes with the smallest  probabilities to be drawn least
 are moved to the tail.
\begin{center}
\fbox{\advance \textwidth by -7em
        \parbox{\textwidth}{\noindent{\bf
 1. Selection of a logarithmic class \boldmath $L_z$}
\begin{enumerate}
\item random number $r\leftarrow \rnd, \quad$ sum $ s\leftarrow 0,\quad $ index
$ i=0$
\item while $s\leq rQ$
  \begin{enumerate}
  \item increase $i\leftarrow i+1$,
  \item add up $s\leftarrow s+\ell_{z_i}$,
  \item if  $i>1$ and $\ell_{z_i}>\ell_{z_{i-1}}$  then\\ \label{bubble}
    exchange the order of the classes $z_i$ and $z_{i-1}$.
  \end{enumerate}
\item return $z=z_i$, i.e~draw $L_{z_i}$.
\end{enumerate}
\advance \textwidth by 7em}}
\end{center}

After $z$ has been selected,
it is easy to  propose a cell ${\vec n}$ by drawing a uniform random
number $u\in \{0,\ldots,\nu_z-1\}$.
We have implemented  each class $L_z$ as an array
 $F_z[0,\ldots,\nu_z-1]$ of $\nu_z=\mid L_z \mid $
elements, each describing the state of  a single cell.
The following subroutine does the von Neumann rejection of a cell in the
previous
selected class $z$.
\begin{center}
  \advance \textwidth by -7em
  \fbox{
    \parbox{\textwidth}{\noindent{\bf 2. Selecting a cell
        \boldmath $\vec n$ in class $L_z$ by von Neumann rejection}
      \begin{enumerate}
      \item propose $u\leftarrow \floor( \nu_z \rnd),\quad \vec n\leftarrow
F_z[u]$,
      \item draw a uniform random number $r\leftarrow \rnd$,
      \item if $ {Q_{\vec n}}>r {2^{z+1}}$ \parbox[t]{20em}{then return $\vec
n$, \\
          else go back to step 1.}
      \end{enumerate}}}
  \advance \textwidth by 7em
\end{center}
Because of the inequality (\ref{the:inequality}) we know lower and upper limits
of the reactivity,  \mbox{$2^{z}\leq Q_{\vec n}<2^{z+1}$}.
Therefore, it  is guaranteed,
that the probability for a rejection
is less than $0.5$.
If we assume that the rates in each class are  distributed uniformly,
we get an acceptance ratio $a=0.75$.
In the cell $\vec n$ the reaction is chosen by a linear selection method,
because the number of possibilities usually is small $O(1)\ldots O(10)$.
\begin{center}
\fbox{\advance \textwidth by -7em
  \parbox{\textwidth}{\noindent
    {\bf 3. Selecting a transition \boldmath $i$ within  cell  \boldmath $\vec
n$},
    \begin{enumerate}
\item  random number $r\leftarrow \rnd,$ sum $s\leftarrow 0$,
\item for all diffusion--transitions $j$:
   \begin{enumerate}
  \item sum $s\leftarrow s+Q^j_{\vec n}$,
  \item if $s \geq r Q_{\vec n}$ \parbox[t]{20em}{then return step $j$,\\
    else next $j$,}
  \end{enumerate}
\item for all chemical reactions $i$,
  \begin{enumerate}
  \item sum $s\leftarrow s+Q^i_{\vec n}$,
  \item if $s \geq r Q_{\vec n}$ \parbox[t]{20em}{then return step $i$,\\
    else next $i$}.
  \end{enumerate}
\end{enumerate}}
\advance \textwidth by 7em}
\end{center}
The changes in all cells and classes involved have to be registered
by  bookkeeping steps
which  require the computation of the reaction--rates $Q^{i}_{\vec n}$, the
local  reactivity of a cell $Q_{\vec n}$, its class reactivity $\ell_{\vec z}$
and the global  reactivity $Q$.
For a diffusion--step concerning a further cell $\vec m$
the bookkeeping has to be done for these cells, too.

\begin{center}
  \advance \textwidth by -7em
  \fbox{
    \parbox{\textwidth}{
      \noindent{\bf 4. Bookkeeping for all cells \boldmath $\vec n,\vec m$
involved}\\
      For all cells $\vec n$, as far as their  rates are concerned do
      \begin{enumerate}
      \item compute the local  rates $Q^i_{\vec n}$,
      \item compute the new cell reactivity $Q_{\vec n}'$ and its new class\\
        $z'\leftarrow\floor(\ld(Q_{\vec n}'))$,
      \item if the class has not changed, i.e.~$z=z'$,\\
        \parbox[b]{7ex}{\parbox[b]{2ex}{\ }then}
        update $\ell_z\leftarrow\ell_z-Q_{\vec n}+Q_{\vec n}'$,\\
        \parbox[b]{7ex}{\parbox[b]{2ex}{\ }else}
        \parbox[t]{25em}{delete cell $\vec n$ in class $z$ and insert it into
$z'$,\\
          including an update  $\ell_z\leftarrow\ell_z-Q_{\vec n}$ and
          $\ell_{z'}\leftarrow\ell_{z'}+Q_{\vec n}'$.}
      \item  update the total reactivity $Q\leftarrow Q-Q_{\vec n}+Q_{\vec n}'$
and
          the cell reactivity $Q_{\vec n}\leftarrow Q_{\vec n}'$,
      \end{enumerate}
      \advance \textwidth by 7em
      }
    }
\end{center}
At this position the algorithm has been presented except for
the  data structure.
Inserting and deleting of cells is somewhat delicate and will be the
subject of the implementation section.

\subsubsection{Implementation of the event handler}
Our approach makes use of a lot of 
sophisticated, non--standard and non--numeric  algorithmic structures.
Because we do not see any way to obtain these results
by usual programming methods, we want to introduce our
data structures in details.
The procedures dealing with these structures have been presented
in the previous section.

Our program has been developed in C  to achieve the highest
portability and speed. We have written the code in an object
oriented style nevertheless using only the ANSI C 2.0 standard.
We implemented the following duties
\begin{enumerate}
\item an initialisation procedure, which serves as a constructor,
  to define the numbers of events and classes   to be managed,
\item a procedure, which can be seen as a
  destructor, to clean up the data structures,
\item a random generator drawing a cell according its reactivity
  due to its contribution following the   algorithm of the section above,
\item a subroutine to insert an event,
\item another subroutine for the deletion of an event.
\end{enumerate}
\subsubsection{Basic data structures}
The most flexible data structures are pointers, thus an event is
described by a pointer to the cell being represented.
Because we do  not want to make any assumption about the cell itself,
we have implemented an event as pointer  to the empty type, i.e. the type
{\tt event} is equivalent to the type {\tt (void *)}. This choice has the
advantage
that it may easily be  converted to a more problem--adapted structure.
The event selection has been kept completely separated
to encapsulate the problem--independent structure of the event handler
from the problem--related structure of the cells and the topology.
\unitlength 1cm
\advance\textwidth by -7em
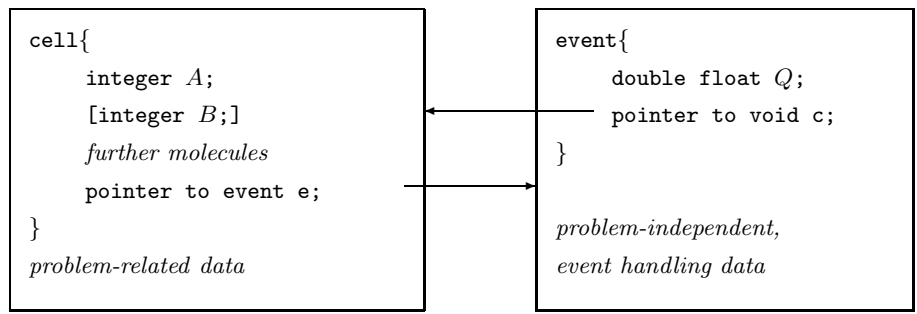
\begin{figure}[ht]
  \begin{center}
    \begin{picture}(10,5)(1,0.5)
      \tt
      \put(0,0.5){\framebox(5.5,4){}}
      \put(0.25,4){cell\{}
      \put(1,3.5){integer $A$;}
      \put(1,3.0){[integer $B$;]}
      \put(1,2.5){\parbox[t]{4cm}{\it further molecules}}
      \put(1,2.0){pointer to event e;}
      \put(5.25,2.15){\vector(1,0){1.75}}
      \put(0.25,1.5){\}}
      \put(0.25,1){\it problem-related data}
      \put(7,0.5){\framebox(5,4){}}
      \put(7.25,4){event\{}
      \put(8,3.5){double float $Q$;}
      \put(8,3.0){pointer to void  c;}
      \put(7.75,3.15){\vector(-1,0){2.25}}
      \put(7.25,2.5){\}}
      \put(7.25,1.5){\it problem-independent,}
      \put(7.25,1){\it event handling data}
    \end{picture}
  \end{center}
  \caption{The cell--event data structure}
\end{figure}
\advance\textwidth by 7em

The {\tt cell}--structure  contains all informations
about the contents of the cell, in our case the integer numbers of
 $A$ or $A$ and $B$ molecules.
Unfortunately we found that the number of events contained in
each class changes rapidly in unpredictable way.
 It is not possible to specify an a priori strategy to determine
the size of a certain class in advance.
We have therefore been forced to write an reorganisation procedure
balancing the memory by dividing it into parts proportional to the size
of each class.

The link  {\tt event  $\to$ cell} is used for
the selection mechanism, i.e~for events caused by a reaction of
$\vec n$. This may be a chemical  reaction within $\vec n$ or
a diffusion--step originating in $\vec n$.
The link from {\tt cell $\to$ event}
is necessary to respond to changes in cell $\vec n$ triggered
by other cells $\vec m$ concerning
$\vec n$, i.e.~some molecule diffusing into $\vec n$.

Since the {\tt event}--structure may be moved by memory reorganisations,
the {\tt cell $\to $ event} link has to be updated every time an event is
moved.
This is achieved by addressing via the link toward the other direction
\mbox{\tt event.cell.e $\leftarrow$ event}.
In general, the structure moving has to inform its related
partner about its new address.

The cells and the events are stored in two separate arrays.
The cells are arranged in a one dimensional array of
size $L^d$ using helical boundary conditions.
The topological neighbourhood of a single cell belonging to
a certain index array
{\tt n}$=\sum_{\delta=0}^{d-1} L^\delta n_\delta $ with $0\leq \mbox{{\tt n}}
\leq L^d-1$
can be addressed by
\begin{equation}
  \mbox{{\tt n}}
  \pm L^\delta \mbox{\ modulo\ } L^d,\qquad 0 \leq \delta \leq d-1.
\end{equation}
 This way the topology  has been  implemented independent
of the integer dimension $d$.
For chemical reactions in fractal dimensions  a more sophisticated structure
is required.
In this case
the neighbourhood of a cell has a  much more complicated topology.

The memory overhead caused by the double link between
event and cell data structures is small and
becomes less important with an increasing number
of reacting species \cite{mydipl}.

\subsubsection{Data structure of the logarithmic classes}
\label{data:structure}
\begin{figure}[p]
  \begin{center}
    \leavevmode
      \tt
      \begin{picture}(15,15)(0,0)
        \put(1,0){\rm classes}
        \put(0.3,5.2){class $z+2$}
        \put(2.8,4.6){\vector(3,-1){2.4}}
        \put(0.8,3.7){$\nu_{z+2}=3$}
        \put(0.8,7.7){$\nu_{z+1}=8$}
        \put(0.8,11.7){$\nu_{z}=5$}
        \put(2.8,8.6){\vector(3,1){2.4}}
        \put(2.8,12.6){\vector(2,1){2.4}}
        \put(0.3,9.2){class $z+1$}
        \put(0.3,13.2){class $z$}
        \multiput(0,2)(0,4){3}{
          \begin{picture}(3,4)(0,0)
            \put(0,0){\framebox(3,4){}}
            \put(0.5,2.7){pointer to}
            \put(0.5,2.2){first of}
            \put(0.5,1.2){events}
          \end{picture}}
        \put(7,0){\rm events}
        \put(12.1,0){\rm cells}
        \multiput(5,2)(0,0.5){25}{
          \begin{picture}(3,0.5)
            \put(0,0){\framebox(5,0.5){}}
          \end{picture}}
        \multiput(11.7,2)(0,0.5){25}{
          \begin{picture}(0,0.5)
            \put(0,0){\framebox(1,0.5){}}
          \end{picture}}
        \put(5.4,14.15){empty}
        \put(5.4,13.65){class \parbox[h]{1cm}{\hfill$z$}, event $1$}
        \put(5.4,13.15){class \parbox[h]{1cm}{\hfill$z$}, event $2$}
        \put(5.4,12.65){class \parbox[h]{1cm}{\hfill$z$}, event $3$}
        \put(5.4,12.15){class \parbox[h]{1cm}{\hfill$z$}, event $4$}
        \put(5.4,11.65){class \parbox[h]{1cm}{\hfill$z$}, event $5$}
        \put(5.4,11.15){empty}
        \put(5.4,10.65){empty}
        \put(5.4,10.15){empty}
        \put(5.4,9.65){empty}
        \put(5.4,9.15){class \parbox[h]{1cm}{\hfill$z+1$}, event $1$}
        \put(5.4,8.65){class \parbox[h]{1cm}{\hfill$z+1$}, event $2$}
        \put(5.4,8.15){class \parbox[h]{1cm}{\hfill$z+1$}, event $3$}
        \put(5.4,7.65){class \parbox[h]{1cm}{\hfill$z+1$}, event $4$}
        \put(5.4,7.15){class \parbox[h]{1cm}{\hfill$z+1$}, event $5$}
        \put(5.4,6.65){class \parbox[h]{1cm}{\hfill$z+1$}, event $6$}
        \put(5.4,6.15){class \parbox[h]{1cm}{\hfill$z+1$}, event $7$}
        \put(5.4,5.65){class \parbox[h]{1cm}{\hfill$z+1$}, event $8$}
        \put(5.4,5.15){empty}
        \put(5.4,4.65){empty}
        \put(5.4,4.15){empty}
        \put(5.4,3.65){class \parbox[h]{1cm}{\hfill$z+2$}, event $1$}
        \put(5.4,3.15){class \parbox[h]{1cm}{\hfill$z+2$}, event $2$}
        \put(5.4,2.65){class \parbox[h]{1cm}{\hfill$z+2$}, event $3$}
        \put(5.4,2.15){empty}
        \put(9.9,14.15){$\emptyset$}
        \put(9.9,13.75){\vector(2,-1){2.0}}
        \put(9.9,13.25){\vector(4,1){2.0}}
        \put(9.9,12.75){\vector(4,-3){2}}
        \put(9.9,12.25){\vector(1,1){2.0}}
        \put(9.9,11.75){\vector(2,-1){2.0}}
        \put(9.9,11.15){$\emptyset$}
        \put(9.9,10.65){$\emptyset$}
        \put(9.9,10.15){$\emptyset$}
        \put(9.9,9.65){$\emptyset$}
        \put(9.9,9.15){\vector(2,-1){2.0}}
        \put(9.9,8.65){\vector(1,-1){2.0}}
        \put(9.9,8.15){\vector(2,1){2.0}}
        \put(9.9,7.65){\vector(1,1){2.0}}
        \put(9.9,7.15){\vector(2,3){2.0}}
        \put(9.9,6.65){\vector(4,1){2.0}}
        \put(9.9,6.15){\vector(4,3){2.0}}
        \put(9.9,5.65){\vector(1,-1){2.0}}
        \put(9.9,5.15){$\emptyset$}
        \put(9.9,4.65){$\emptyset$}
        \put(9.9,4.15){$\emptyset$}
        \put(9.9,3.65){\vector(4,3){2.0}}
        \put(9.9,3.15){\vector(1,0){2.0}}
        \put(9.9,2.65){\vector(1,1){2.0}}
        \put(9.9,2.15){$\emptyset$}
      \end{picture}
  \end{center}
  \caption{\label{fig:structure}{Data structure of classes,
      events and cells. At first a class is selected, then an event related to
a cell,
      and finally a reaction in a cell.
      The class structure points  to the first of its events.
      Among the event sub--arrays of different classes empty places are left.}}
\end{figure}
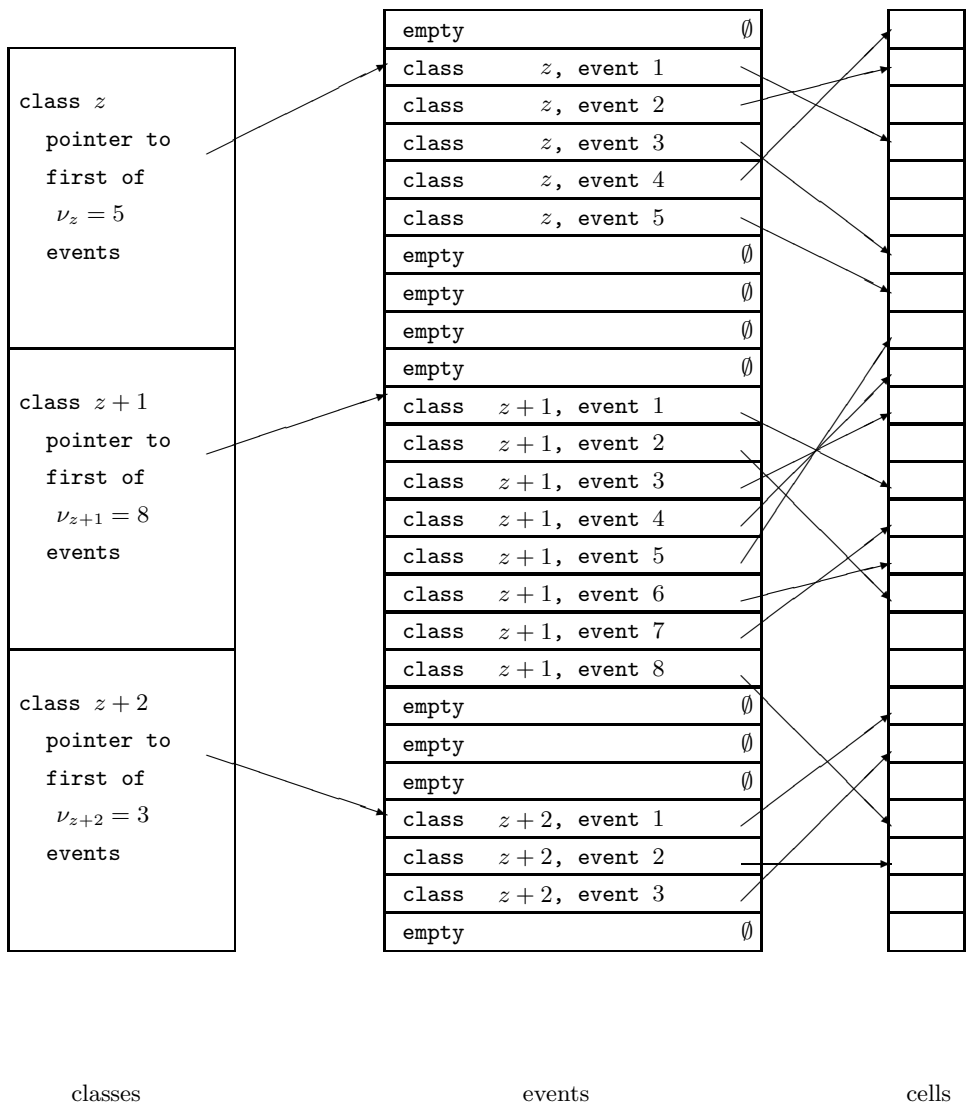

The array of events has a substructure appropriate to the
contents of the logarithmic classes.
For each class  a descriptor contains all related information.
\begin{center}
\begin{tabular}[h]{llll}
  \\class\{&\\
  &{pointer to next class $L_{z_{i+1}}$;}
  &{\it sequence of classes as  a linked list}
  \\
  &{integer $z_i$;}
  &{\it the power of 2 related to the class}
  \\
  &{\it derived information}
  &{\it for speedup only}
  \\
  &{double float $\ell_z$;}
  &{\it class reactivity}
  \\
  &{integer $\nu_z$;}
  &{\it number of events in class}
  \\
  &{$F_z[]$ as a pointer to $e$ ;}
  &{\it the first element of the class}\\
  \}\\[1ex]
  \end{tabular}
\end{center}
Within each class the events are stored in a one dimensional array,
which is to contain no empty places.
The classes are realized as structures containing several elements,
as shown above.
The sequence for the linear selection is realized
by a linked list, so there is a pointer to the next class.
The integer $z$ contains the power of two managed by
the structure.  For speedup, some related frequently used  information
as $2^z,2^{z+1}$ etc.~also may be stored in this structure.
The number  $\nu_z$  describes the variable size of
the array of events $F_z[]$, which is implemented as a pointer to its first
element.
At this point, we use the ability of the programming language
C to  access entries by indirect or relative addressing.
The complete data structure including
the links to the cells is shown by figure \ref{fig:structure}.
\advance\textwidth by -7em
\advance\textwidth by 7em

If an event is inserted, it is simply  stored  at the end of
the (sub)array $F_z[]$, after that $\nu_z\leftarrow \nu_z+1$ is increased.
If an event $i$ is deleted,
gaps in the array of events $F_z[]$ must be avoided
to keep the von Neumann rejection efficient.
Therefore the last event of the sub-array
has to be moved in the gap, $F_z[i]\leftarrow F_z[\nu_z]$, before
 decreasing $\nu_z\leftarrow \nu_z -1$.
Because there is a cell structure pointing to this event,
its cell has to be informed that its
related event has moved.

Since  there is no a--priori strategy to estimate the
strong fluctuations in the size of the classes, we run into
the problems of collisions in memory.
 If the array of events $F_z[]$ of  a single
class $L_z$ is going to overwrite
 the array of events   $F_{z+1}[]$
of its successor, it is necessary to reorganise the division
of the memory occupied by the array of events.
This is done by crunching and re-expanding all classes
proportional to their actual  memory consumption.
\advance\textwidth by -7em
\begin{figure}[hpt]
  \begin{center}
    \leavevmode
   \mytwothirdfpicture{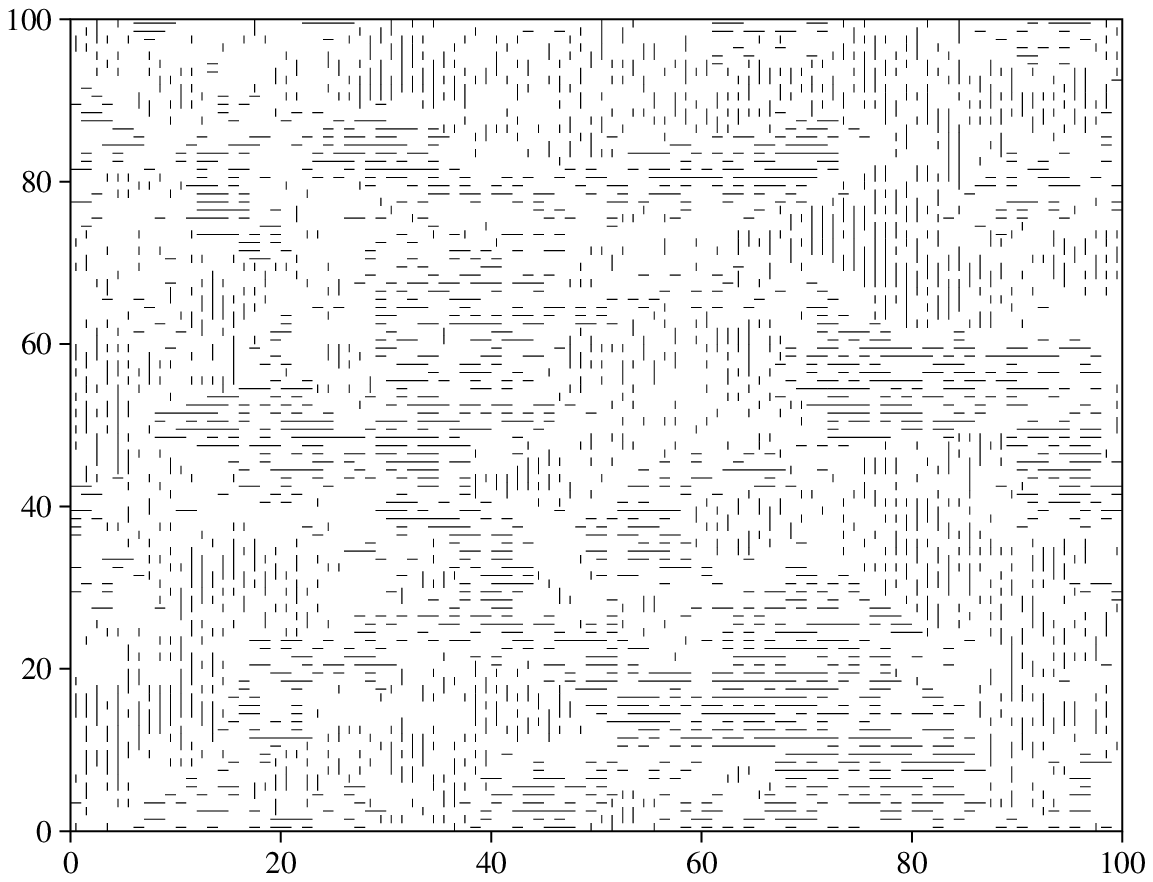}{
     \put(-1.4,5){$y$}
     \put(4.8,-1){$x$}
     }
  \end{center}
  \caption{\label{DomPic3}Formation of domains in the diffusion--controlled
reaction
    \mbox{$A+B\to \emptyset$}       for a volume of $100\times 100$ cells.
    The symbol "$\mid$" denotes cells containing one or more
    $A$ molecules while the symbol "$-$" denotes $B$.
    }
\end{figure}
\advance\textwidth by 7em

\subsection{Implementation of the reactions studied}
According to  our approach to separate the event handling from
the chemistry within a cell, most of the work is done
once the event handler has been implemented.
The cells are realized as structured  variables according to
section (\ref{data:structure}).
Thus adapting the algorithm to any specific RDS does not require
more than the implementation for drawing a reaction within a cell
to select the  specific reaction.
In our case  we have to choose
\begin{enumerate}
\item a volume $\Omega$,
\item the number of cells $L^d$ defining a discretisation
  $ \omega={\Omega}/{L^d}$ and $h=\omega^{{1}/{d}}$,
\item a hopping rate $\lambda_{\alpha\vec n\vec m}={D_\alpha}/{h^2}$,
\item and a local reaction--constant
  $\lambda_k =\omega{k}/{\omega^2} ={k}/{\omega} $.
\end{enumerate}
Another procedure has to be provided to do the related changes and
to compute the reaction--rates.
According to  (\ref{vantHoff}) and (\ref{diffusion})
the reaction-- and diffusion--rates
of cell $\vec n$ are given by
\begin{eqnarray}
  \Lambda_{\vec n}&=&\parbox{10em}{$\lambda_k X_{A\vec n}(X_{A\vec
n}-1)$}\mbox{or}\\
  \Lambda_{\vec n}&=&\parbox{10em}{$\lambda_k X_{A\vec n}X_{B\vec
n}$}\mbox{and}\\
  \Lambda_{\alpha\vec n}&=&\parbox{10em}{$X_{\alpha\vec n} \lambda_{\alpha\vec
n\vec m}$}
  \mbox{with $\alpha=A,B$}.
\end{eqnarray}
\advance\textwidth by -7em
\begin{figure}[hpt]
  \begin{center}
    \leavevmode
       \mytwothirdfpicture{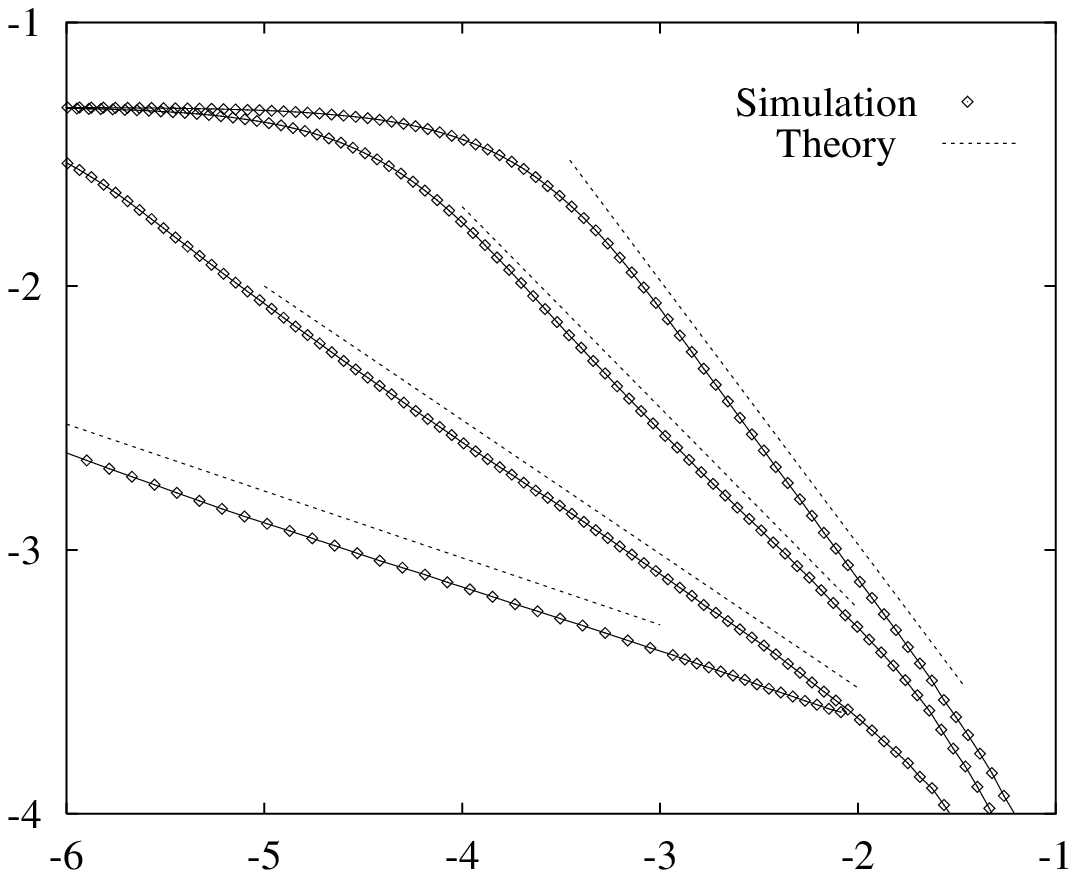}
             {
               \put(0,-1.2){\parbox{6.6666cm}{\centering time  $\quad\lg_{10}
t$}}
               \put(0,10.3){\parbox{6.6666cm}{\centering density}}
               \put(-3.5,5){$\lg_{10} n_{A,B}(t)$}
              \put(1.2,2.8){$d=1$}       \put(1.1,5.3){$d=2$}
              \put(2.6,6.8){$d=3$}       \put(6.3, 7){$d=4$}
             }
  \end{center}
  \caption{\label{fig:QAB}Simulation of $A+B\to\emptyset$ for
                        diffusion-controlled reactions in the
                        dimensions $d=1,2,3,4$, extension
                        $L^d=3\cdot 10^4,\, 400^2,\, 50^3,\, 25^4$, $\triangle
n=0$}
\end{figure}
\begin{figure}[t]
  \begin{center}
    \leavevmode
       \mytwothirdfpicture{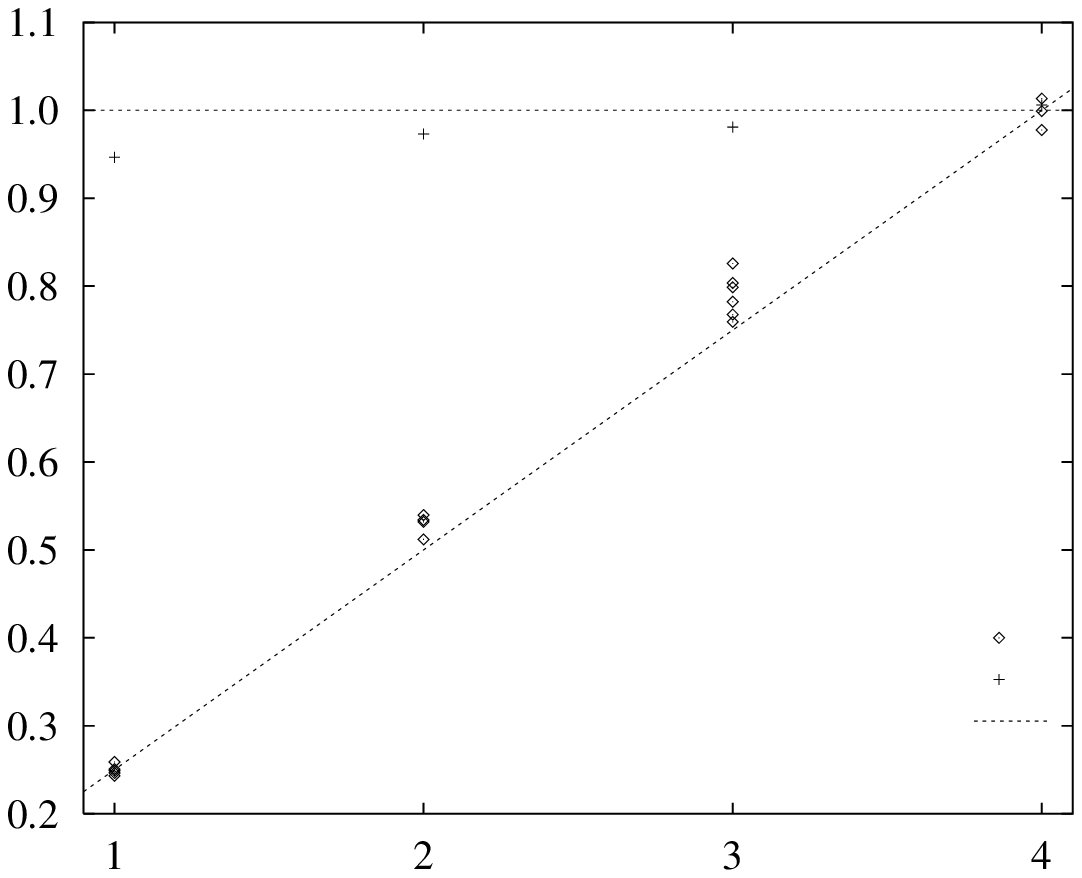}
                  { \put(3.8,-1.2){dimension $d$}
                   \put(0,10.2){\parbox{6.667cm}{\centering $A+B\to \emptyset$,
                                                         exponents $\gamma$ }}
                   \put(2.55,2.1){ diffusion-controlled $k>D$}
                   \put(2.55,1.55){ reaction-controlled $k<D$}
                   \put(2.55,1.0){ theoretical exponents}
                   \put(7.2,5){\ $\gamma =\frac{d}{4}$}
                   \put(7.1,5.1){\line(-2,1){1}}
                   \put(4,9.2){\ $\gamma=1$}
                   \put(3.9,9.3){\line(-1,-1){0.4}}
                   \put(-2,5){$\gamma$}
                  }
  \caption{\label{fig:AB}
    {Simulation of $A+B\to\emptyset$ for
      diffusion--controlled reactions in the
      dimensions $d=1,2,3,4$, numerical values in table
(\protect{\ref{tab:AB}})}}
  \end{center}
\end{figure}
\begin{figure}[hpt]
  \begin{center}
    \leavevmode
    \mytwothirdfpicture{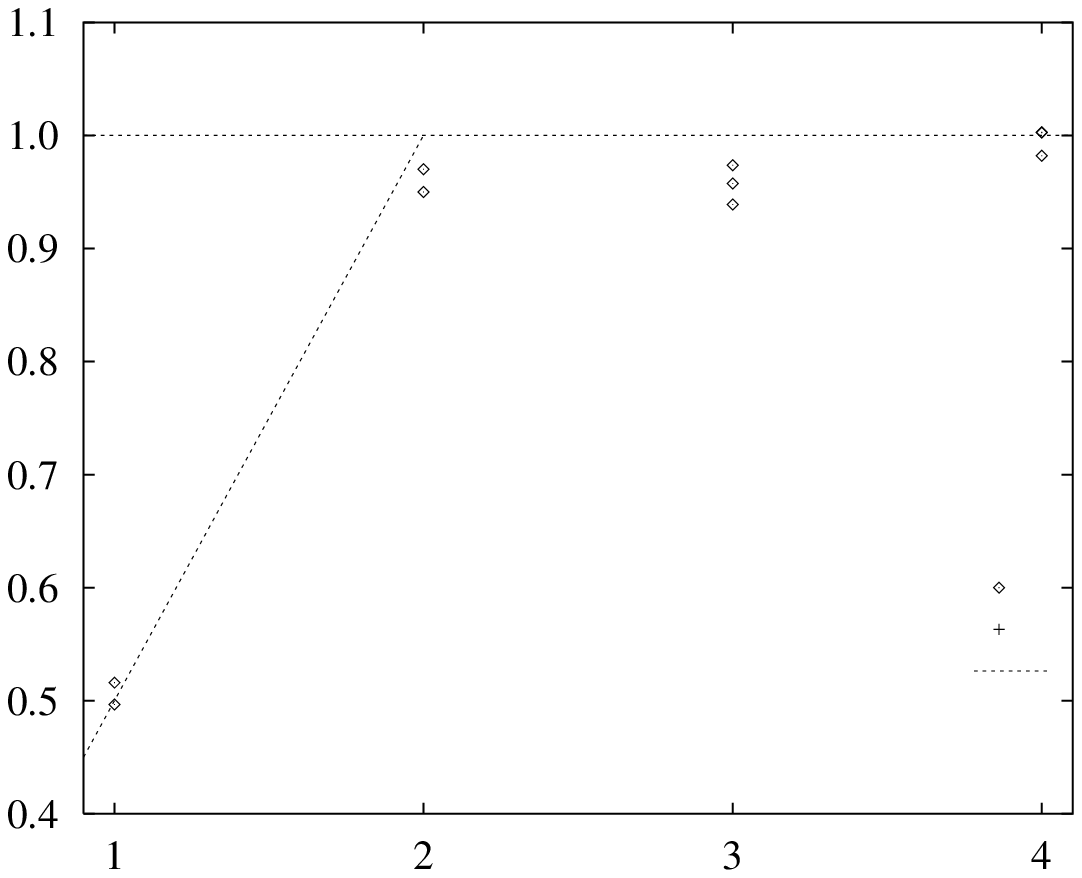}
                  { \put(3.8,-1.2){dimension}
                   \put(0, 10.2){\parbox{6.6667cm}{\centering
                        $A+A\to \emptyset$, exponents $\beta$ }}
                   \put(3.8,-1.2){dimension $d$}
                   \put(2.8,2.8){diffusion-controlled $k>D$}
                   \put(2.8,2.25){reaction-controlled $k<D$}
                   \put(2.8,1.7){theoretical exponents}
                   \put(3.4,4.7){\ $\beta =\frac{d}{2}$}
                   \put(3.3,4.8){\line(-2,1){1.1}}
                   \put(6,9){\ $\beta =1$}
                   \put(5.9,9.1){\line(-1,-1){0.5}}
                   \put(-2,5){$\beta$}
                  }
  \caption{\label{GrDiPlAA}\label{fig:AA}
              {Exponents of the algebraic decay of $A+A\to\emptyset$ for
               different dimensions and several
               reaction--probabilities $k=10^s$, numerical values in
               table (\protect{\ref{tab:AA}})}}
\end{center}
\end{figure}
\advance\textwidth by -7em

\section{Simulations in several dimensions}
The aim of  our simulations is to test the accuracy of the
scaling ansatz.
We want  to examine both limits of the reaction--diffusion--system.
On the one hand, there is the reaction--controlled limit,
where  the dynamics is determined by the reaction--constant
 \mbox{${D}/{\Omega^{2/d}}\gg k$}, expecting  mean--field behaviour
$n_\alpha(t)\propto t^{-1}$,
on the other hand  we examine  diffusion--controlled
limit ${D}/{\Omega^{2/d}}\ll k$,
where the long time annihilation is delayed by the necessity
of transport.

Defining length and time scales we have chosen $\Omega=1$ and
 $D=D_A=D_B=1$.  This way the reaction constant $k$ is fixed,
too.  Since $k$ now has the meaning of a rate in the case of one
molecule per unit volume, $D$ and $k$ get inverse times
 $[D]=[k]=[1/t]$.

\subsection{The system $A+B\to 0$}
Figure (\ref{DomPic3}) shows a snapshot of the
formation of domains,
figure (\ref{fig:QAB}) shows the time dependent density
$n_\alpha (t)$ for the diffusion--controlled
 $A+B\to \emptyset$ reaction.
The  logarithmic plot
compares the results of our simulations  to straight lines $t^{-d/4}$
predicted by theory.
The size of the system obviously is sufficiently large,
thus one run in each dimension may be considered as self--averaging.
The figures (\ref{fig:AB}) and (\ref{fig:AA}) show the
exponents of the algebraic annihilation
for both  diffusion-- ($D\ll  k$) and  reaction-- ($D\gg  k$) controlled
limits.
In our simulations  these inequalities  are realized by
a ratio of at least 9 orders of magnitude for $k\gg D=1$.
To be sure to be sufficiently close to the limit $k/D\to \infty$
the rate $k$ has been increased until the exponents measured by linear
regression have not changed any more.

Obviously, the scaling theory for the $A+B\to \emptyset$ reaction
is within the statistical errors of a few percent.
\advance\textwidth by -7em
\begin{table}[ht]
  \begin{center}
    \leavevmode
    \begin{tabular}{r|r|r|r|r}
      $d$ & 1 & 2 & 3 & 4\\\hline
      & $ 0.250_9 $ & $0.534_9 $ & $0.826_4 $ & $0.978_6$ \\
      $\gamma_s\qquad\ $ & $0.256_9$    & $ 0.512_9   $ & $ 0.803_5   $ &
$0.999_7$\\
      & $  $ & $  $ & $ 0.799_5  $ & $$\\
      & & & $0.782_6$&\\\hline
      & $ 0.249_{12} $ & $0.532_{11}  $ & $ 0.768_9  $ & $ 1.013_9 $\\
      $\gamma_{\infty}\qquad$ & $ 0.243_{12}$  & $ 0.540_{11} $ & $ 0.759_9  $
& $ $\\
      & $ 0.246_{12} $ & $ $ & $    $ & $ $
    \end{tabular}
  \caption{\label{tab:AB}
    { Numerical results of $A+B\to\emptyset$, see figure
(\protect{\ref{fig:AB}}).
      The values are presented as
      $\gamma_s$, where the subscript $s$ denotes the order of magnitude
      of $k=10^s$ with $\gamma_{\infty}$ representing the
      best approach to the diffusion--controlled
      limit,  extension  $L^d=30\,000^1,\, 400^2,\, 50^3,\, 25^4, \quad
n_A(0)=n_B(0)$)}}
  \end{center}
\end{table}
\begin{table}[ht]
  \begin{center}
    \leavevmode
    \begin{tabular}{r|r|r|r|r}
      $d$ & 1 & 2 & 3 & 4\\\hline
      $\beta_s\qquad$  & $ 0.5160_{11}$ & $ $ & $$ & $ 0.9821_7 $\\\hline
               & $ 0.4966_{12}$ & $ 0.9686_9 $ & $ 0.9390_9$ & $ 1.0028_9 $\\
               $\beta_\infty\qquad$  & $ $ & $ 0.9504_9 $ & $ 0.9736_9$ &
$1.0025_9$\\
               &&&$ 0.9575_9$&
    \end{tabular}
  \end{center}
  \caption{\label{tab:AA} Numerical results of $A+A\to\emptyset$, see figure
(\protect{\ref{fig:AA}}).
    As in the previous table the subscript $s$ denotes the
               order of magnitude of the reaction constant.
               Note,  that in $d=d_c=2$
               logarithmic corrections have been performed.}
\end{table}
\advance\textwidth by 7em
\subsection{The system $A+A\to 0$}
For the $A+A\to \emptyset$ reaction we observe
logarithmic behaviour in its critical dimension, as predicted
by \cite{bra:leb}.
This is not a multiparticle effect and results from
two dimensional Smoluchowsky theory for only two particles
The $\frac{\log(t)}{t}$ has divided by $\log(t)$
to reproduce the correct $t^{-1}$ scaling.
The results have been obtained by $k=10^s=10^9\ldots 10^{12}$,
thus we are convinced, that our simulations
give an excellent approximation to the  diffusion--controlled limit.
The simulations have been done
with some $10^7$ molecules.
For every  case studied, we have performed a
finite size  control by doubling the size $L$ until
we have not find any  influence of the
finite volume. Furthermore, we could neglect corrections,
even if in $d=4$ where we have been limited to a hypercube
of linear extension $L=25$, what may be a consequence of $d_c=4$.

\section{Conclusion}
Thus within statistical errors of $1\ldots 2\% $
and  an uncertainty due to the method of linear regression of
the same order of magnitude,
the tables  (\ref{tab:AB}) and (\ref{tab:AA})
show, that scaling theory describes the $A+A\to \emptyset $
and $A+B\to \emptyset $ correctly, including the important
cases $A+A\to \emptyset $ and the logarithmic corrections in
 $d=2$, where we
found  an annihilation according to $n_A(t)\propto \frac{\log t}{t} $.

Thus, we have shown that for the simplest reaction--diffusion--systems,
in which the influence of the finite number of molecules in a single cell
must be taken into account, the algorithm is able
to cover the whole field from diffusion-- to  reaction--controlled
systems.
A detailed study of the crossover of both
regimes which  determines the
crossover time  depending on the size of the system and the
diffusion constants will follow.
For the simplest reactions, the theoretical results have been
reproduced nearly exactly.

Our approach follows a strategy opposite to the common cellular
automaton philosophy.
Instead of simulating a lot of simple diffusion--steps represented by integer
operations, we try to find a length scale below which diffusion may be
neglected.
On this scale one single hopping process may replace a large number
of infinitesimal automaton lattice gas steps.

The price to pay is the implementation of a  complex event handler
to maintain the exact Markoff--probabilities.
{}From our experience we are pleased that the total overhead produced
by the event handler is less than a factor of 3--4 per Markoff--event
compared to reaction--systems without diffusion.
The requirement of diffusion being much faster than chemical
reactions causes an additional overhead by a factor of 5--10.

The typical speed of a common Ultrix or Alpha 80 Mc RISC
workstations leads to some $10^4-10^5 $
Markoff--operations per second.
For some $10^7$ annihilation processes we need some 10 minutes
to a few hours.

We claim our  algorithm to be superior to any lattice gas
simulation, because we are not restricted to one molecule
per cell.
For $k\to \infty $ the lattice gas is included as limit,
for $k\to 0$ the simulation degenerates to the homogeneous case.
An upper limit for the size $\omega$ of the reaction cells is
given by the comparison of reaction and diffusion--time--scales.


The computing time for more complex RDS will increase linearly
with the number of reactions.
In our simple cases the memory overhead also is about a factor of 2,
and can be compared to two additional components of the RDS.
However, after we have implemented an event handler
which is completely independent of physical  problem and the
underlying topology, further simulations  of RDS are
realized with little effort.
The event handler can be easily integrated
into any program,
because the only necessary adaptions  are  the
selection of the reactions in a single cell.

\subsection{Future extensions}
Due to the generality of the method, we see a lot of possible applications.

\subsubsection{Obvious generalisations}
Further extensions are possible. If its neighbourhood
influences the reactions of a cell, i.e.~if we want to
describe surface dynamics or the fluctuations of a field,
the general method keeps unchanged.
Only a slight modification of
the update of the neighbour cells must be taken into account.

The simplest assumption, the chemical reactions being independent of $\vec n$,
may be completed by reactions depending on the single cell or a cell and its
topological neighbourhood. With regard to these conditions  we only have
to  supplement further indices if necessary.
The transition probabilities,  which do not only depend on the single
cell $\vec n$, but also on its neighbourhood may be denoted by the symbol
 $M_{\vec m\vec m'\vec m''\ldots}$.
This way we can indicate the  possibility
that a  step depends on the contents
of more than one single cell.
For those types of events the symbol  $M$ is in contrast
to  $\Lambda$, which shall be left to local rates.
Therefore $M^i_{\vec n\vec m\vec m'\ldots}$ depends on
the cells $\vec n,\vec m,\vec m'\ldots $,
whereas $\Lambda^i_{\vec n}$ is restricted to $\vec n$.
If we provide  $\Lambda_{\vec n\vec m\vec m'\ldots }^i$
with additional indices, this symbol serves as a denotation
for a transition { changing the contents of several cells}
$\vec n,\vec m,\vec m'$
 with a rate
{depending only on the single cell $\vec n$}.
This is the manner we have   described diffusion in Gillespie's algorithm
by a reaction changing the contents of two cells,
whose rate $\Lambda_{\alpha\vec n\vec m}$ only  depends on $\vec n$ alone.

To handle a reaction as a Markoff--event by the event handler,
it is always assigned to its first cell specified.
Thus $\Lambda^i_{\vec m\ldots}$ and $M^i_{\vec n\vec m\ldots}$
both are  assigned to cell $\vec n$.
\begin{eqnarray}\nonumber
\label{Q:rate:general}
  Q_{\vec n}&=&\Lambda_{\vec n}+M_{\vec n}
  \\
  &=&
  \sum_i \Lambda^i_{\vec n}+
  \sum_i \Lambda^i_{\vec n\vec m}+\cdots+
  \sum_i M^i_{\vec n}
  +\sum_{i} M_{\vec n\vec m}^{i}
  +\sum_{i} M_{\vec n\vec m\vec m'}^{i}+\cdots.
\end{eqnarray}

The difference of the dependencies on one or more
cells is unimportant for the principal selection of a reaction
step.
Again an arbitrary event is unambiguously denoted by $i$ and $\vec n$
and its rate $Q^i_{\vec n}$.
More precisely, this symbol now unifies the description of
 $  \Lambda^{i}_{\vec n},  \Lambda^{i}_{\vec n\vec m},M^{i}_{\vec n\vec m\ldots
}$.
For a transition touching further cells $\vec m,\vec m'$
the bookkeeping had to be done for these cells, too.

For RDS the case of a reaction depending on several cells
does not attend,
we have presented this generalised formalism to show,
that even more complex dynamics could be handled.
This way it is possible to introduce events depending on gradients
into the framework of chemical reactions.
This   opens an outlook not only to the simulation
of surface dynamics, but points out to new approach to
field theoretical models on computers.

We want to proceed to simulate
more complex behaviour, with a variety of possible
chemical reactions.

\subsubsection{Automatically generated code}
Therefore, we are writing a  code generator,
generating C source code by a script
to automatise the process
from the problem definition to the program running.
As we have seen, more complex reactions
are difficult to implement because for any practical purpose,
any efficient code expands
from one line definition to one page C source code.
This will be the subject of a future paper.

We expect even much more interesting stochastic  effects
for more complex systems or in more complicated topologies.
Simulations become essential,
because  by the lack of a simple scaling ansatz we may
lose control over the
fluctuations in a complex system,
especially if the numbers of molecules in an individual cell are small.


\appendix{:\ Scaling approach for $A+A\to \emptyset$}
\noindent
\label{appendix:ab}
In the diffusion--controlled limit,
the mean field behaviour cannot correctly describe the role
of fluctuations.
The approach of the scaling theory is to derive
arguments connecting fluctuations to a length scale
$\xi$.
The length $\xi$  is related to a typical time scale $t_\xi$
by standard diffusion--scaling $t_\xi=\xi^2/D$.

Kang and Redner  have  argued \cite{kang:redner},
that the initial mean distance $\xi=a_{NN}$ of a $A$ molecule
to its next neighbour is proportional to $n(0)^{-1/d}$.
Therefore, the time-scale $\tau_{NN}$ to pass this distance is scaled
by the standard diffusion--scaling of space and time according to
\begin{equation}
  \tau_{NN}\cong \frac{a^2_{NN}}{D}=n(0)^{-\frac{2}{d}}.
\end{equation}
The density is scaled by a function  as $g(x)$
\begin{equation}
  n(t)=n(0)\cdot g\left(\frac{t}{\tau_{NN}}\right).
\end{equation}
with the asymptotic properties
\begin{eqnarray}
  \parbox[t]{10ex}{$x\ll 1$} g(x)&= &1,\\
  \parbox[t]{10ex}{$x\gg 1$}g(x)&\propto & x^{-\beta}.
\end{eqnarray}
To compute the scaling exponent $\beta$, we need
the additional assumption,
that the long time behaviour is  independent  of the initial
concentration, thus we obtain
\begin{equation}
    (Dt)^{\beta}n(t)=\mbox{constant}.
\end{equation}
In $d_c$
 the scaling and the mean--field exponents meet, thus
\begin{equation}
  \beta=\frac{d}{2},\qquad d_c=2.
\end{equation}

\appendix{:\ Scaling approach for $A+B\to \emptyset$}
The dynamics of this system are governed by domains,
which are defined  as  local areas where either the $A$ or
the $B$ species dominates. A typical picture obtained by simulations
in $d=2$ is shown in Fig. \ref{DomPic3}.

Because $A$ and $B$
annihilate as pairs, the difference
\begin{equation}
\label{difference}
  \Delta n=n_A(t)-n_B(t)=n_A(0)-n_B(0)=\mbox{constant}
\end{equation}
is conserved.
In a volume $\zeta^d$  of linear extension $\zeta$
the initial number of molecules of each species\
may vary  due to  fluctuations at time $t=0$
\begin{equation}
\label{xi:def}
  N_\alpha(0)=\langle N_\alpha\rangle \pm \sqrt{\langle N_\alpha\rangle }=
  n_\alpha(0)\zeta^d\pm \sqrt{n_\alpha(0)\zeta^d}.
\end{equation}
Some species may
locally be in majority even it is global in minority
i.e.~$N_A>N_B$ even if $n_A<n_B$.
Thus by fluctuations local inversions are possible.
The largest possible volume $\xi^d$, in which an inversion
may occur,  is estimated by equating
the numbers of $A$ and $B$ molecules
\begin{equation}
\label{diffAB}
  0=N_A(0)-N_B(0)=
  \left|\sqrt{N_A}+\sqrt{N_B} \right|.
\end{equation}
Inserting (\ref{xi:def}) we solve  (\ref{diffAB})  obtaining
the length scale
\begin{equation}
  \xi=\left|\sqrt{n_A(0)}-\sqrt{n_B(0)}\right|^{-\frac{2}{d}}.
\end{equation}
By diffusion--scaling  $\xi$ is related to the time required to cross
a domain
\begin{equation}
  t_{\xi}=\frac{\xi^2}{D},
\end{equation}
assuming both diffusion--constants being equal, i.e.~$D=D_A=D_B$.
The scaling ansatz
\begin{equation}
  n_\alpha(t)=C_\alpha t^{-\gamma} f_\alpha\left(\frac{t}{\tau_\xi}\right)
\end{equation}
requires the existence of scaling functions $f_\alpha(x)$,
whose exponential decay dominates the algebraic decay
 $t^{-\gamma}$ for $t>t_\xi$.
For short times the scaling functions are assumed to be constant,
thus
\begin{eqnarray}
  \parbox[t]{10ex}{$x\ll 1$} f_\alpha(x)&= &1,\\
  \parbox[t]{10ex}{$x\gg 1$} f(x)&\propto & \exp(-c x).
\end{eqnarray}
Inserting $f_\alpha$   into (\ref{difference})
we get
\begin{equation}
\label{cabcb}
  C_A\cdot f_A\left(\frac{t}{t_{\xi}}\right)-
  C_B\cdot f_B\left(\frac{t}{t_{\xi}}\right)=
  \frac{1}{2}\left|\sqrt{n_A(0)}-\sqrt{n_B(0)}\right|\cdot
t^{-\frac{d}{4}}_{\xi}t^\gamma.
\end{equation}
The left side of the equation only depends
on the ratio ${t}/{t_{\xi}}$, therefore this statement
yields  for the right side of this equation, too,
thus leading to  the algebraic exponent
\begin{equation}
  \gamma=\frac{d}{4},\qquad d_c=4.
\end{equation}
Although in the limit $n_A(0)\to n_A(0)$ the factor
$\left|\sqrt{n_A(0)}-\sqrt{n_B(0)}\right|\to 0$
in equation (\ref{cabcb}) disappears, the
argumentation is not affected and the scaling keeps valid
\cite{kang:redner,kang:redner2}.

\end{document}